\documentclass[conference]{IEEEtran}
\IEEEoverridecommandlockouts
\usepackage[T1]{fontenc}
\usepackage{amsmath}
\usepackage{amssymb}
\usepackage[pdftex]{graphicx}
\usepackage{xspace}
\usepackage{url}
\usepackage{lmodern}
\usepackage{subfig}
\usepackage{xcolor}

\newtheorem{definition}{Definition}

\newcommand*\OK{$\checkmark$}

\begin{document}

\title{The Long Road to Computational \\Location Privacy: A Survey*
\thanks{\textsuperscript{*}This paper is to appear in IEEE Communications Surveys \& Tutorials.}
}

\author{\IEEEauthorblockN{Vincent Primault}
	\IEEEauthorblockA{University College London\\
		United Kingdom\\
		v.primault@ucl.ac.uk}
	\and
	\IEEEauthorblockN{Antoine Boutet}
	\IEEEauthorblockA{INSA Lyon, Inria, CITI\\
		Villeurbanne, France\\
		antoine.boutet@insa-lyon.fr}
	\and
	\IEEEauthorblockN{Sonia Ben Mokhtar}
	\IEEEauthorblockA{INSA Lyon, CNRS, LIRIS\\
		Villeurbanne, France\\
		sonia.benmokhtar@insa-lyon.fr}
	\and
	\IEEEauthorblockN{Lionel Brinoe}
	\IEEEauthorblockA{INSA Lyon, CNRS, LIRIS\\
		Villeurbanne, France\\
		lionel.brunie@insa-lyon.fr}
}

\maketitle

\begin{abstract}
The widespread adoption of continuously connected smartphones and tablets developed the usage of mobile applications, among which many use location to provide geolocated services.
These services provide new prospects for users: getting directions to work in the morning, leaving a check-in at a restaurant at noon and checking next day's weather in the evening are possible right from any mobile device embedding a GPS chip.
In these location-based applications, the user's location is sent to a server, which uses them to provide contextual and personalised answers.
However, nothing prevents the latter from gathering, analysing and possibly sharing the collected information, which opens the door to many privacy threats.
Indeed, mobility data can reveal sensitive information about users, among which one's home, work place or even religious and political preferences.
For this reason, many privacy-preserving mechanisms have been proposed these last years to enhance location privacy while using geolocated services.
This article surveys and organises contributions in this area from classical building blocks to the most recent developments of privacy threats and location privacy-preserving mechanisms.
We divide the protection mechanisms between online and offline use cases, and organise them into six categories depending on the nature of their algorithm.
Moreover, this article surveys the evaluation metrics used to assess protection mechanisms in terms of privacy, utility and performance.
Finally, open challenges and new directions to address the problem of computational location privacy are pointed out and discussed.
\end{abstract}

\section{Introduction}
\label{sec:introduction}

More and more users are carrying a handheld device such as a smartphone or a tablet and using it to access a wide variety of services on the go.
Using these services, users can consult their bank accounts' balance from anywhere, book a table in a restaurant at any moment, track the status of their flight in real time or get a notification when their friends are nearby.
This is possible thanks to the wide range of sensors available on these devices, giving them access to some knowledge about their environment.
They are able to determine their location in real time and then use it to interact with geolocated services, often called \textit{Location-Based Services} (LBSs for short).
These services provide a contextual and personalised information depending on the current user's location.
A multitude of LBSs have emerged these last years.
We give here a non-exhaustive list of common usages that have been enabled by the rise of LBSs.

\begin{itemize}
    \item \textit{Directions \& navigation applications:}
    These services allow users to get directions to (almost) any destination, and then to navigate towards it by simply following spoken instructions.
    Location data is used to provide real-time directions, recalculated as the user is moving.
    Well-known players here include Google Maps~\cite{gmaps} and Waze~\cite{waze}.
    \item \textit{Weather applications:}
    These services provide current weather conditions as well as forecasts.
    Location data is used to give the user relevant information for the city he is currently located in.
    Yahoo! Weather~\cite{yahooweather} is an application providing such a service on Android and iOS.
    \item \textit{Venue finders:}
    These services give users information about interesting places in their vicinity.
    Most of the time, they include recommendations based on the experience of other users.
    Location data is used to show only places in immediate user's neighbourhood.
    Foursquare~\cite{foursquare} and Yelp~\cite{yelp} are two applications helping to find such interesting places, with an added social dimension.
    \item \textit{Social games:}
    These services turn any urban walk into an ever-changing game, where each new place becomes a new playground.
    Location data is used to make the game evolve depending on the user's city and his immediate surroundings, sometimes allowing to compete with nearby other users.
    Examples of such games are Pokemon GO~\cite{pokemongo} and City Domination~\cite{citydomination}.
    \item \textit{Crowd-sensing applications:}
    These services enable participatory sensing, where a crowd of users use their smartphones to monitor their environment and share their results through an LBS server.
    Crowd-sensing benefits to a large variety of domains such as smart cities (e.g., the traffic monitoring application Nericell~\cite{Mohan08}) or health monitoring (e.g., PEIR~\cite{Mun09}).
    APISENSE~\cite{Haderer13} and Funf~\cite{Aharony11} are two applications allowing to run crowd-sensing campaigns.
\end{itemize}

%%% Figure: LPPMs use cases.
\begin{figure*}[ht]
  \centering
  \includegraphics[width=0.86\textwidth]{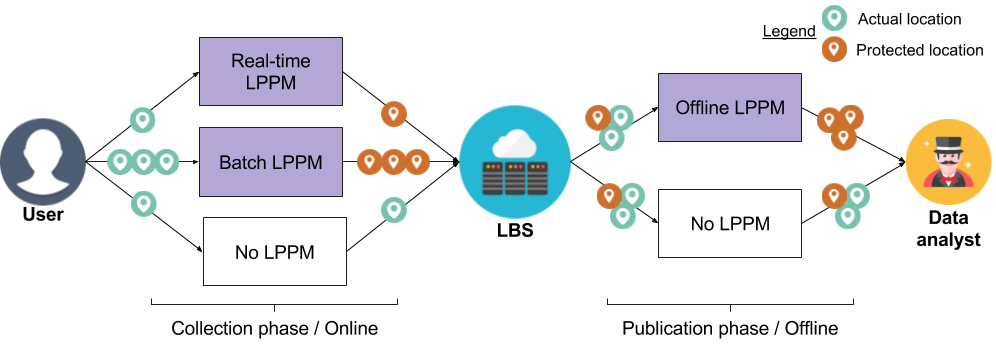}
  \caption{Depending on the use case (i.e., real-time, batch, or offline), an LPPM can operate on location data during the collection or the publication phase.}
  \label{fig:intro:lppms}
\end{figure*}

Whatever their exact nature these services require users to disclose their location in order to make the application working as expected.
This location disclosure nevertheless causes loss of control from users on their privacy and consequently allows LBSs to store the mobility and all places users are visiting over time.
The sequence of all locations known to belong to a single user, along with the time at which the user was seen at each location, is called a \textit{mobility trace}.
Users are often not aware of the quantity of sensitive knowledge that can be inferred from their mobility traces.
Analysing mobility traces of users can reveal their \textit{points of interest}~\cite{Gambs11}, which are meaningful places such as home or work. 
It can also reveal the other users they frequently meet~\cite{Sharad14}, or lead to predicting their future mobility~\cite{Sadilek12}.
By combining mobility traces with semantic information~\cite{Krumm13}, it is also possible to infer the actual user's activity (e.g., working, shopping, watching a film) or its transportation mode if the user is on the move.
Besides the continuous tracking of a user's activities, points of interest can lead to leak even more sensitive information such as religious, political or sexual beliefs if one regularly goes to the headquarters of a political party, a worship place or a lesbian bar, respectively.
As an example, it is possible to find out which taxi drivers are Muslim by correlating the time at which they are in pause with mandatory prayer times~\cite{Franceschi15}.

Needless to say, this large amount of mobility data is a gold mine for many companies willing to learn more information about users.
The market related to LBSs is indeed enormous: the total revenue of the US-only LBS industry was already estimated to \$75 billion in 2012~\cite{Henttu12}.
This high value of the mobility data leads many applications and companies to commercially exploit the collected data for analysis, profiling, marketing, behavioural targeting, or simply to sell this information to external parties.
Moreover, the location of users is tracked and collected by many mobile applications with and without their consent~\cite{Enck10, Achara2013, DBLP:journals/popets/EskandariKAOC17, ZDG+15, Almuhimedi:2015:YLS:2702123.2702210}, which aggravates the privacy threats related to sharing mobility data, voluntarily or not.

To mitigate these privacy problems, many \textit{location privacy protection mechanisms} (LPPMs for short, or just \textit{protection mechanisms} in this article) have been proposed in the literature.
Their goal is to protect location privacy of users while still allowing them to enjoy geolocated services.
There is a rich literature about existing LPPMs.
Some of them are rather generic and can adapt to a lot of situations while others are very specific to a single use case.
LPPMs rely on a wide array of techniques, ranging from data perturbation (e.g.,~\cite{Andres13,Ghinita07,Jiang13}) to data encryption (e.g.,~\cite{Zhong07,Mascetti11,Popa11}), and including fake data generation (e.g.,~\cite{Quercia11,Pelekis11,Kido05}).
In this survey, we distinguish between three classes of use cases for LPPMs as illustrated in Figure~\ref{fig:intro:lppms}.

In \textit{real-time use cases}, users query an LBS and expect an immediate answer.
We include in this category the usage of navigation applications, weather applications, venue finders and social games.
The main challenge for real-time LPPMs (e.g.,~\cite{Andres13,Ghinita07,Popa11}) is that they only have at their disposal actual and historical locations; they obviously do not know the future state of the system.

\textit{Offline use cases} come into play once an LBS has collected mobility data and wants to publish it, whether it is for commercial or non-profit purposes (e.g., sharing mobility data with a marketing company or to release a dataset as open data).
Instead of protecting locations on-the-fly, offline LPPMs (e.g.,~\cite{Abul10,Mir13,Gramaglia15}) protect whole mobility datasets at once, possibly leveraging the knowledge of the behaviour of all users in the system to apply more efficient and subtle schemes.

In \textit{batch use cases}, users regularly send their data to an LBS (e.g., every hour) and expect it to publish back aggregated results.
This use case is typically adopted by crowd-sensing applications~\cite{Mousa15} and is a middle-ground situation between real-time and offline use cases.
Batch LPPMs differ from real-time LPPMs in that they are less sensitive to latency, and they send more data at once.
They also differ from offline LPPMs because they do not have the global knowledge.

\begin{figure*}[!t]
  \centering
  \includegraphics[width=12.4cm]{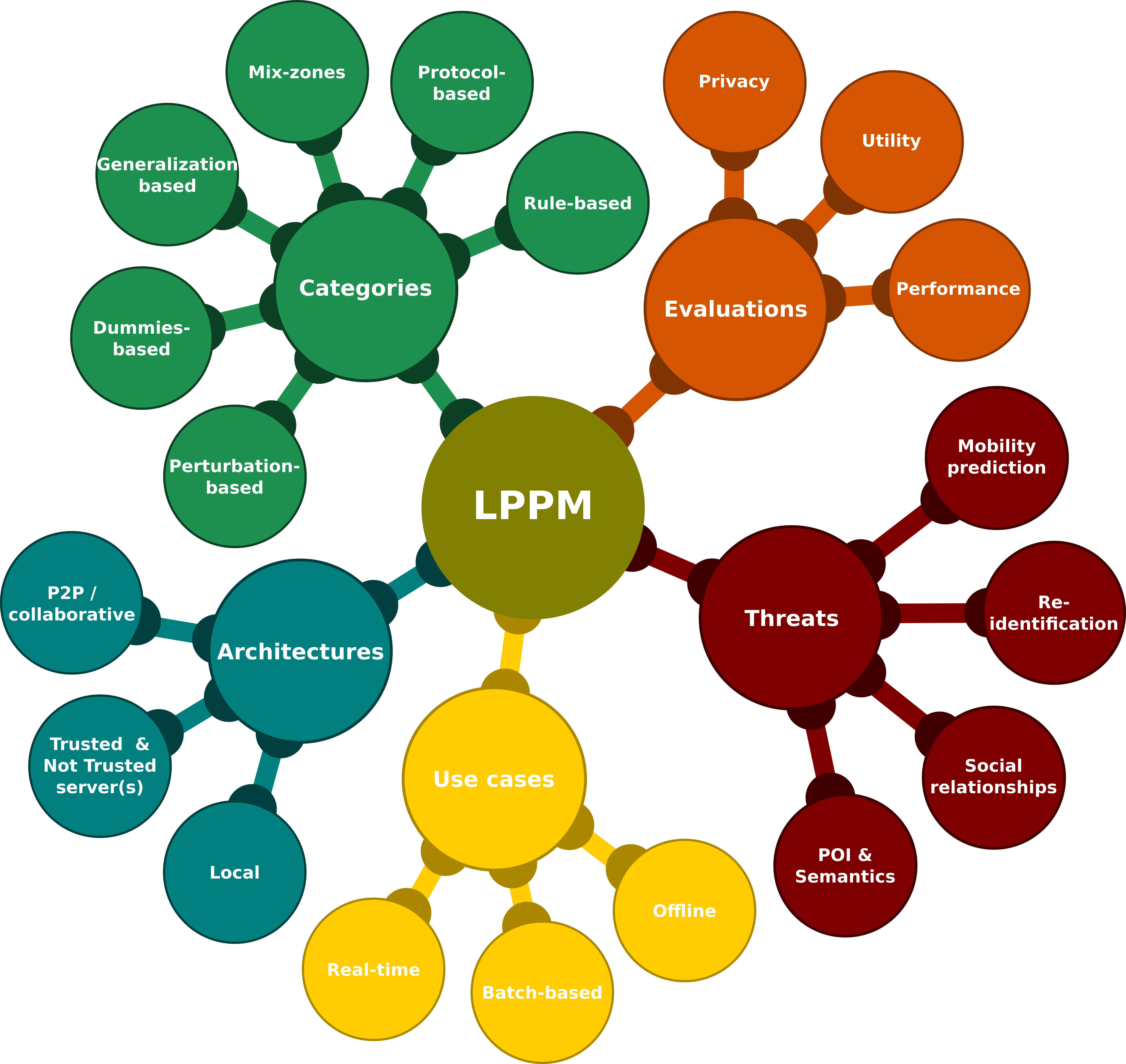}
  \caption{Our survey reviews practical threats associated with location disclosure and surveys state-of-the-art LPPMs over different use cases (i.e., protecting data in real-time, operating by batches, or offline). LPPMs are structured in six categories and can adopt different architectures. We also detail their evaluation in terms of privacy and utility, as well as performance.}
  \label{fig:overview}
\end{figure*}

Figure~\ref{fig:intro:lppms} depicts an overview of the three aforementioned uses cases as well as the involved entities.
As shown, we distinguish between two phases when using an LPPM: what happens \textit{online} during the \textit{collection}, i.e., between a user and an LBS, and what happens \textit{offline}, during the \textit{publication}, i.e., between an LBS and an analyst.
Depending on their nature, LBSs fall either in the real-time or batch use cases.
Furthermore, offline use cases appear as soon as one of these LBSs is willing to publish and to protect the gathered mobility data.

Protecting mobility data with an LPPM obviously improves the privacy but also impacts the quality of the resulting data.
To evaluate LPPMs and to compare them together, researchers have proposed a large variety of metrics.
These metrics can be divided in three categories.
\textit{Privacy metrics} quantify the level of privacy a user can expect while using a given LPPM.
One popular way to evaluate privacy is to evaluate the effect of a privacy attack while being protected by an LPPM (e.g.,~\cite{Shokri11}).
\textit{Utility metrics} measure the usefulness (also called quality of service) that can still be obtained while using an LPPM, which largely depends on the targeted LBS and the considered use case.
There is an inherent trade-off between privacy and utility.
Indeed, if no mobility data is sent, privacy is perfectly preserved, while the utility is null.
Conversely, sending unprotected data results in a perfect utility at the cost of no privacy.
Finally, \textit{performance metrics} measure the algorithm efficiency or the cost of a given LPPM.
Typical performance metrics are the execution time, the ability to scale or the tolerance to faults.
Performance metrics are orthogonal and do not participate to the privacy and utility trade-off, but still are important because they impact the usability of LPPMs.

With the advent of pervasive computing offering new possibilities for tracking and collecting the location and the mobility of users, computational location privacy has become essential.
We are only interested in computational location privacy, i.e., in threats and countermeasures performed by algorithms.
This survey is not related to non-computational threats, which would for example come from a manual inspection and reasoning on mobility data, performed by a human brain.
Computational location privacy has already been reviewed in several surveys.
While both Krumm~\cite{Krumm09b} and Shin et al.~\cite{Shin12} published general surveys, 
Terrovitis~\cite{Terrovitis11} and Wernke et al.~\cite{Wernke14} followed an approach focused on location privacy attacks, and Chow et al.~\cite{Trajectory-privacy} focus on online LPPMs.
However, only few of these articles address the evaluation of protection mechanisms, and when it is actually discussed, only privacy is considered without the complementary utility and performance metrics.
Moreover, previous surveys often focus either on the online or the offline scenario but not both.
In the context of Cognitive Radio Networks (CRNs), Grissa and al.~\cite{7898414} provide a comprehensive survey that investigates the various location privacy risks and threats as well, as countermeasures that have been proposed in the literature to cope with these location privacy issues.
Lastly, Cottrill~\cite{urisa} uses a multidisciplinary approach to discuss location privacy and to ensure that the protection of private information is directly addressed from each of the relevant standpoints of law, policy, and technology.

In this survey, we provide an up-to-date vision over computational location privacy, including recent works like differentially private approaches~\cite{Andres13} or privacy-by-design LBS architectures~\cite{Guha12}.
Figure~\ref{fig:overview} depicts the different areas covered by our survey.
We review practical privacy attacks on mobility data, and survey state-of-the-art protection mechanisms.
As presented in Figure~\ref{fig:overview}, we organise LPPMs into three use cases (i.e., real-time, batch, or offline), and into six categories depending on the kind of algorithm (i.e., mix-zones, generalization-based, dummies-based, perturbation-based, protocol-based and rule-based).
We furthermore discuss the evaluation of LPPMs with privacy, utility, and performance metrics, and report how the presented protection mechanisms have been evaluated by their authors. 
We also consider their architecture and associated impacts (e.g., added latency or integrability).
Lastly, from the lessons we learned from our experience, we discuss open challenges and emerging visions in computation location privacy. 

In this survey, we do not consider protection mechanisms only using anonymity, also called pseudonymity. 
These solutions consist in removing the link between an individual and its data by using a pseudonym instead of his real identity.
Indeed, only relying on pseudonymization to protect data is not enough and this practice has resulted in several well-known privacy breaches these last years. (e.g., the identification of the governor of Massachusetts in an "anonymised" health dataset~\cite{Sweeney02} or the re-identification of users from AOL web search logs~\cite{aol} or a Netflix dataset~\cite{Narayanan08}).

% Paper organisation.
The remainder of this survey is structured as follows.
We present in Section~\ref{sec:threats} the practical threats associated with location disclosure, before reviewing the evaluation metrics used to assess LPPMs in Section~\ref{sec:evaluation}.
We present the different architectures adopted by LPPMs in Section~\ref{sec:archi}.
We then survey state-of-the-art protection mechanisms in Section~\ref{sec:lppms}.
We finally present some open challenges in Section~\ref{sec:challenges} before concluding in Section~\ref{sec:conclusion}.

\section{Privacy threats}%
\label{sec:threats}

Although LBSs potentially provide useful services, users are not always aware of the risks associated with the disclosure of their location during their daily life.
For instance, the goal of the website \textit{Please Rob Me}~\cite{pleaserobme} is to "raise awareness about over-sharing".
They use geolocated tweets to infer whether a user is at home or not, and hence if the way is free for potential thieves.
In this section, we present the most important considered adversary models and the main practical threats for a user related to the exploitation of her mobility traces.

%%%%%%%%%%%%%%%%%%%%%%%%%%%%%%%%%%%%%%%%%%%%
\subsection{Adversary models}
\label{sec:threats:model}
%%%%%%%%%%%%%%%%%%%%%%%%%%%%%%%%%%%%%%%%%%%%

An adversary is anyone who can have an access to mobility data of one or several users.
According to Figure~\ref{fig:intro:lppms}, the adversary can be either the LBS itself or a data scientist who gets a dataset after it was published.
In most of the cases, the adversary is considered to be \textit{honest-but-curious}~\cite{Goldreich:2003:CCP:966037.966044}.
This means that the data scientist or the LBS behaves correctly (i.e., it provides the expected service) but it may exploit in all possible ways the information it receives.
In particular, we assume that the adversary has access to any database containing additional information about the semantic of places or the associated activities, the topology of the road network, details about the public transportation lines (e.g., map, timetables), and so on. 
In addition, we also assume that the adversary may be able to collect external knowledge about each user in the system, modelled as a set of past mobility data.
For instance, it can be used to run state-of-the-art re-identification attacks in order to re-associate the received data to a known user, or to predict future mobility. 
Moreover, it is also assumed that the LBS may identify that the client is relying on a LPPM while communicating with it.
In some cases, it is further assumed that the adversary may collude with the proxy (i.e., typically when the proxy is not trusted) or different nodes in the system (e.g., using a Sybil attack~\cite{Douceur2002}) in order to learn more information about the users.

%%%%%%%%%%%%%%%%%%%%%%%%%%%%%%%%%%%%%%%%%%%%
\subsection{Points of interest \& semantics}
\label{sec:threats:semantics}
\begin{figure}[ht]
  \centering
  \includegraphics[width=0.24\textwidth]{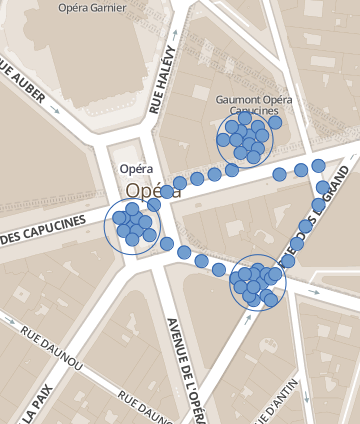}
  \caption{Three POIs have been extracted from this mobility trace by using a clustering algorithm.}
  \label{fig:pois}
\end{figure}

%%%%%%%%%%%%%%%%%%%%%%%%%%%%%%%%%%%%%%%%%%%%
\textit{Points of interest} (POIs for short) are spatially delimited places where users spend some time.
POIs can be home or work places, but also a swimming pool, a school or a cinema for instance.
But they can also be even more sensitive places like a religious monument where a user regularly goes, the headquarters of a political party he is involved in or a hospital he is cured in.
They can be extracted from mobility traces quite easily by using simple clustering algorithms like the ones presented in~\cite{Zhou04,Hariharan04}.
Figure~\ref{fig:pois} shows an example of the behaviour of such an algorithm designed to extract POIs.
Using APIs such as the Google Places~\cite{palceAPI} for instance can also provide details about each POI such as its precise address, the associated activity or the shops located at this location.

Gambs et al.~\cite{Gambs11} made an attack on a dataset containing mobility traces of taxi drivers in the San Francisco Bay Area.
By finding points where the taxi's GPS sensor was off for a long period of time (e.g. 2 hours), they were able to infer POIs of the drivers.
In some cases, they were able to locate a home and even to confirm it by using a satellite view of the area showing the presence of a yellow cab parked in front of the supposed driver's home.
It was possible to infer a plausible home in a small neighbourhood for 20 out of 90 mobility traces analysed.
Deneau~\cite{Franceschi15} created a visualisation tool to analyse the active and inactive periods of taxi drivers over the day.
By correlating their time of inactivity with the five times of prayer per day observed by practising Muslims, it was possible to identify drivers that could be Muslims.

The new development of machine learning brings huge security risks on location privacy. 
For instance, machine learning has showed his effectiveness to learn the semantics of some place.
Krumm introduced \textit{Placer}~\cite{Krumm13}, a system using machine learning to automatically label places into 14 categories (home, work, shopping, transportation, place of worship, etc.).
Author reported an overall accuracy of 73~\%, mostly thanks to home and work places which are the easiest ones to label because this is where people spend most of their time.
Riederer et al. developed \textit{FindYou}~\cite{Riederer16}, which aims to raise user awareness on the privacy issues surrounding the collection and use of location data.
FindYou allow users to import their own location data from popular social networks and audit them. 
By leveraging the knowledge provided by the US Census Bureau and unsupervised machine learning methods, this personal location privacy auditing tool 
provides prediction on the home place of users as well as their age, income and ethnicity.
Huguenin et al.~\cite{Huguenin17} leveraged machine learning to infer the motivation (e.g., "Inform about activity", "Share mood" or "Wish people to join me") behind check-ins of users on Foursquare.
They achieved up to 63~\% of accuracy when predicting a coarse-grained motivation.

%%%%%%%%%%%%%%%%%%%%%%%%%%%%%%%%%
\subsection{Social relationships}
%%%%%%%%%%%%%%%%%%%%%%%%%%%%%%%%%
With several mobility traces, it is possible to compare them and infer relationships between users.
The idea is very simple: if two (or more) persons spend some time within the same area at the same moment, they are likely to be related by some social link.
Bilogrevic et al.~\cite{Bilogrevic13} studied this threat by using malicious Wi-Fi access points deployed on the EPFL campus (in Switzerland) that were able to locate devices communicating with them.
By setting appropriate thresholds, they could detect meetings between people. 
Then, they used training data to get a characterisation of the social link between students, thus finding out if they were classmates, friends or other depending on the place where they met and the time they spent together.
With the optimal experimental parameters, the authors experimentally obtained a true positives rate of 60~\% and a false positives rate of 20~\%.
Similarly, Wang et al.~\cite{Wang17} studied how access points could infer social relationships between users.
They designed a decision tree classifying relationships by using the interaction time, inferred activity type (i.e., home, work, leisure) and physical closeness of people.
Their classifier exhibits a success rate above 85~\%.

%%%%%%%%%%%%%%%%%%%%%%%%%%%%%%
\subsection{Re-identification}
%%%%%%%%%%%%%%%%%%%%%%%%%%%%%%
Mobility traces can ultimately lead to re-identifying physical users, i.e., associating an identity to each trace.
Krumm~\cite{Krumm07} used two months of mobility traces and tried to infer users' home address with various heuristics.
By using white pages, it was possible in some cases to associate a person to a mobility trace.
The most accurate heuristic gave 9 correct re-identifications out of 172 drivers.
Gambs et al.~\cite{Gambs14} proposed a re-identification approach based on mobility Markov chains.
They are used to model mobility patterns of users, more specifically the transitions between POIs.
They designed different distance measures to quantify the similarity between two Markov chains and used them to re-identify users.
They achieved up to 45~\% of good matching, which was significantly better than other state-of-the-art attacks.
Tockar~\cite{Tockar14} showed it was possible to stalk at celebrities by using a taxi trips dataset and some specifically crafted queries.
By extracting drop-off addresses of people frequently spending their night in a club, and by using Google and Facebook to get more information about these addresses, he was able to pinpoint certain individuals with a high probability.
Pyrgelis~\cite{Pyrgelis18} et al. studied the feasibility of membership inference attacks on aggregate mobility datasets (i.e., datasets featuring how many users where within specific regions during specific periods of time).
They modeled the problem as a game, and trained a machine learning classifier on prior knowledge and used it to infer whether particular individuals were part of a mobility dataset.
Powerful attackers reached an Area Under Curve up to 0.83 or even 1.0, depending on how much background knowledge they have at their disposal.

These results are mainly possible due to the high degree of uniqueness of human mobility.
Indeed, De Montjoye et al.~\cite{DeMontjoye13} showed that only four randomly chosen spatio-temporal points are sufficient to almost uniquely identify a user hidden among the crowd.
It means that the mobility of every individual acts like a fingerprint, even among a large number of users (they used a dataset containing 1.5 million users).
In the same manner, Golle et al.~\cite{Golle09} studied the uniqueness of the home/work pair.
When revealing the census district where people live and work, it is possible to uniquely identify 5\% of them, and for 40\% of them, they are only hidden among 9 other people.
If the census block where they live and work is disclosed, nearly all people can be uniquely identified.
Zang et al.~\cite{Zang11} improved the previous study by considering top-N locations of a large set of call data records of a US nation-wide cell operator.
With the most precise location that can be grabbed from these records, which is the sector where the person stands, it is possible to uniquely identify 35\% of the users by using their top-two locations and 85\% of them by using their top-three locations.
We also demonstrated the highly unique nature of mobility traces from different sensors~\cite{Boutet16}.
Among other observations, authors noticed that the temporal dimension is as discriminative as the spatial dimension and that the uniqueness degree of mobility traces is user-dependent.
They also pointed out that it is still possible to re-identify users with a high success rate using an appropriate attack even if the data was protected by classical LPPMs.
Manousakas et al.~\cite{Manousakas18} showed it might be even worse, as even if removing all spatial and temporal information from the graph of visited places, the topology of the latter graph is itself often uniquely identifying.
They evaluated their approach with a dataset of 1500 users and found out that about 60~\% of the users are uniquely identifiable from their top-10 locations, and this percentage increases to 93~\% in the case of a directed graph.
With a directed graph, 19 locations are needed to uniquely identify each of the 1500 users.

Recently, Wang et al.~\cite{Wang18} explored the discrepancies between the theory and practice of re-identification attacks.
They leveraged a large ground-truth dataset containing 2 million users, and two smaller external datasets collected over the same population to match against the former.
They first evaluated seven state-of-the-art algorithms, showing that with those datasets, the best performing ones only achieved a re-identification rate of 20~\%, which is far from the theoretical bound announced by De Montjoye~\cite{DeMontjoye13} and others.
By further analysing the results, it appeared that there was large spatio-temporal mismatches, whose effect was underestimated.
Finally, they proposed four new re-identification attacks addressing the previously mentioned concerns, improving the re-identification of 17~\% as compared to previous state-of-the-art algorithms.

%%%%%%%%%%%%%%%%%%%%%%%%%%%%%%%%%%%%%%%
\subsection{Future mobility prediction}
%%%%%%%%%%%%%%%%%%%%%%%%%%%%%%%%%%%%%%%
Knowing past mobility of a user can help to model his habits and hence allow to predict where he will be in a future time.
Noulas et al.~\cite{Noulas12} focused on Foursquare check-ins.
They extracted features from users mobility and used machine learning algorithms to predict places where users were likely to leave the next check-in.
Precision was maximal during morning and at noon, when they achieved an accuracy of 65~\%.
It was harder to predict the next check-in in the night, during which accuracy dropped to 50~\%.
Sadilek and Krumm~\cite{Sadilek12} proposed \textit{Far Out}, a system to predict the location of a user in the long term, i.e., in a far away future date and within a time window of one hour. 
They leveraged Fourier analysis and principal component analysis to extract repetitive patterns from mobility data.
These patterns were associated to a week day and a hour in the day.
Their system featured an accuracy in their predictions ranging from 77~\% to 93~\%.
Another threat coming from the analysis of mobility traces is the inference of users' mobility patterns.
Gambs et al.~\cite{Gambs12} modelled movement habits of people by using Markov chains.
Each frequent POI becomes a state in the chain and a probability is assigned to each possible transition.
They split the day into several temporal slices and differentiated between week days and week-ends.
With a dataset spanning over a sufficient period of time, it became possible to predict future users' movements.
Agir et al.\cite{Agir16} studied the prediction of the next check-in of users by using Bayesian networks.
They evaluated their approach with data of 1065 users, collected from tweets that have been generated from Foursquare.
Across 6 major cities, the median accuracy when predicting the next check-in is between 100 and 150 metres.
The authors also studied the impact of semantic information on this kind of attack.
They found out that by incorporating semantics, the median accuracy of their predictions decreased by 10 to 115 metres.

\section{Evaluating LPPMs}
\label{sec:evaluation}

Unfortunately, there is no standard to evaluate and compare LPPMs.
To the best of our knowledge, only the work of Shokri et al.~\cite{Shokri11} focuses on the evaluation of LPPMs.
Although it is only interested in quantifying privacy, it defines solid foundations towards building a complete evaluation methodology.
In this section, we review the different evaluation metrics used in the literature to assess LPPMs in a quantitative manner.
We start by introducing two classical privacy notions in Section~\ref{sec:background:notions}.
We then group and present evaluation metrics through three complementary categories, namely privacy metrics in Section~\ref{sec:background:privacy}, 
utility metrics in Section~\ref{sec:background:utility} and performance metrics in Section~\ref{sec:background:performance}. 
We discuss the inherent trade-off between privacy and utility in Section~\ref{sec:evaluation:tradeoff}.
We conclude this section by surveying mobility datasets commonly used to conduct practical evaluations in Section~\ref{sec:evaluation:datasets}.

%%%%%%%%%%%%%%%%%%%%%%%%%%%%%%%%%%%%%%%%%%%%%
\subsection{Privacy models}
\label{sec:background:notions}
%%%%%%%%%%%%%%%%%%%%%%%%%%%%%%%%%%%%%%%%%%%%%
Two general privacy models have emerged and have been widely adopted by the community, and are still the foundation for most of subsequent works~\cite{DBLP:series/synthesis/2016Domingo-Ferrer}.
Those models propose generic privacy guarantees that were originally not specific to location privacy, but have been later successfully applied to location privacy.
In this subsection (and only this one), the concept of dataset is not limited to mobility datasets but to generic datasets, i.e., a list of records with attributes.

%%%%%%%%%%%%%%%%%%%%%%%%%%%%%%%%%%%%%%%%%%%%%
\subsubsection{$k$-anonymity}
\label{sec:background:kanonymity}
%%%%%%%%%%%%%%%%%%%%%%%%%%%%%%%%%%%%%%%%%%%%%
The model of $k$-anonymity has been introduced by Sweeney in 2002~\cite{Sweeney02}.
The idea is to prevent one to uniquely identify individuals from a small subset of their attributes, called a \textit{quasi-identifier}.
The subset of attributes to protect, which is not part of the quasi-identifier, form the \textit{sensitive attributes}.
For instance, within medical records, the birth date, sex and zip code triplet is a quasi-identifier that is enough to uniquely identify some individuals, while the disease is a sensitive attribute.
$k$-anonymity states that to be protected, a user must be indistinguishable among at least $k-1$ other users.
To achieve that, all $k$ indistinguishable users must have the same values for all attributes forming their quasi-identifier.
This makes them look similar and forms what is called an \textit{anonymity group}.
Therefore, the probability of an attacker without external knowledge to re-identify someone among $k$ similar users is at most $1/k$.

\begin{definition}
  Let $d$ be a sequence of records with $n$ attributes $a_1,...,a_n$ and $Q_d = \{a_i,...,a_j\} \subseteq \{a_1,...,a_n\}$ be the quasi-identifier associated with $d$.
  Let $d_k$ be the $k$-th record of $d$ and $r[Q_d]$ the projection of record $r \in d$ on $Q_d$, i.e., the $|Q_d|$-tuple formed of values for only the attributes of $Q_d$ in $r$.
  $d$ is said to satisfy $k$-anonymity if and only if each unique sequence of values in the quasi-identifier appears with at least $k$ occurrences in $d$, or formally:
  \[ \forall s \in \{ r[Q_d] \, | \, r \in d \}, | \{ i \in \mathbb{N} \, | \, d_i[Q_d] = s \} | \geq k \]
\end{definition}

\begin{table}[ht]
  \centering
  \caption{A dataset with $k$-anonymity where $k=2$.}
  \label{tab:background:notions:kanonymity}
  \begin{tabular}{|l|l|l|l|}
    \hline
    Birth & Sex & Zip & Disease \\
    \hline
    1970 & M & 0247 & Migraine \\
    1970 & M & 0247 & Chest pain \\
    1970 & F & 0247 & Asthma \\
    1970 & F & 0247 & Migraine \\
    1970 & F & 0247 & Asthma \\
    1969 & M & 0232 & Appendicitis \\
    1969 & M & 0232 & Appendicitis \\
  \hline
  \end{tabular}
\vspace{2mm}
\end{table}

For example, Table~\ref{tab:background:notions:kanonymity} shows a sample of a medical dataset exposing a $k$-anonymity guarantee, where the quasi-identifier is $\{Birth,Sex,Zip\}$ and the sensitive attributes are $\{Disease\}$, for $k=2$. 
Here, there are three unique $\{Birth,Sex,Zip\}$ triplets, i.e., $\langle 1970, M, 0247 \rangle$, $\langle 1970, F, 0247 \rangle$ and $\langle 1969, M, 0232 \rangle$.
For each of those triplets, there are respectively two, three and two different records.
Consequently, there is a minimum of two different records for each triplet of values taken by the quasi-identifier: this table guarantees 2-anonymity.
This way, knowing the birth year, sex and zip code of some individual should not leak his disease, as there is at least one other person with the same quasi-identifier.

\begin{table*}[ht]
  \centering
\caption{Two datasets differing on one single element.}
\label{tab:background:notions:dp}
  \begin{tabular}{cc}
    \begin{tabular}[b]{|l|c|}
      \hline
      Name & Has chronic migraines \\
      \hline
      Agatha & True \\
      Anna & False \\
      John & True \\
      Mark & False \\
      Mary & False \\
      \hline
    \end{tabular} & 
	\begin{tabular}[b]{|l|c|}
      \hline
      Name & Has chronic migraines \\
      \hline
      Agatha & True \\
      Anna & False \\
      Joe & True \\
      John & True \\
      Mark & False \\
      Mary & False \\
      \hline
    \end{tabular} \vspace{3mm}\\

	(a) Dataset $D_1$, without Joe. & (b) Dataset $D_2$, with Joe. \\
  \end{tabular}
\end{table*}

However, despite providing 2-anonymity, there is a problem in Table~\ref{tab:background:notions:kanonymity} for male patients born in 1969 and living in the area with 0232 zip code (i.e., the last two records).
Indeed, they share the same value for their sensitive attribute (i.e., they have the same disease), which leaves them unprotected.
This concern has been addressed by the introduction of $\ell$-diversity~\cite{Machanavajjhala07}.
It extends $k$-anonymity by additionally enforcing that within anonymity groups, there should be at least $\ell$ well-represented values.
More precisely, it enforces a particular distribution of values for sensitive attributes across each anonymity group.
This well-represented notion is formally defined in three different ways in~\cite{Machanavajjhala07}.
The simplest one is called distinct $\ell$-diversity and states that there must be at least $\ell$ distinct values for each sensitive field for each anonymity group.

$t$-closeness~\cite{Li07} is a further extension of $\ell$-diversity % proposed by Liu et al. 
Instead of just guaranteeing a good representation of sensitive values, this approach enforces that the distribution of every sensitive attribute inside anonymity groups must be the same than the distribution of this attribute in the whole dataset, modulo a threshold $t$.

%%%%%%%%%%%%%%%%%%%%%%%%%%%%%%%%%%%%%%%%%%%%%
\subsubsection{Differential privacy}
\label{sec:background:dp}
%%%%%%%%%%%%%%%%%%%%%%%%%%%%%%%%%%%%%%%%%%%%%
Differential privacy is a more recent model introduced by Dwork~\cite{Dwork06} defining a formal and provable privacy guarantee.
The idea is that an aggregate result computed over a dataset should be almost the same whether or not a single element is present inside the dataset.
In other words, the addition or removal of one single element shall not change significantly the probability of any outcome of an aggregate function.
Unlike $k$-anonymity, the differential privacy definition is not affected by the external knowledge an attacker may have.

\begin{definition}
  Let $\epsilon \in \mathbb{R}^{+*}$ and $\mathcal{K}$ be a randomized function that takes a dataset as input.
  Let $image(\mathcal{K})$ be the image of $\mathcal{K}$.
  $\mathcal{K}$ gives $\epsilon$-differential privacy if for all datasets $D_1$ and $D_2$ differing on at most one element, and for all $S \subseteq image(\mathcal{K})$,
  \[\Pr[\mathcal{K}(D_1) \in S] \leq e^{\epsilon} \times \Pr[\mathcal{K}(D_2) \in S]\]
\end{definition}

For example, Table~\ref{tab:background:notions:dp} shows two versions of a sample dataset listing whether individuals are subject to chronic migraines.
Let us suppose that an analyst has access to these two datasets, and to a query $Q$ that takes a dataset as input and returns the number of persons having chronic  migraines.
By computing $Q(D_2)-Q(D_1) = 3-2 = 1$, our curious analyst can infer that Joe is indeed subject to chronic migraines.

Several methods have been proposed to practically achieve differential privacy.
We present one of them, called the Laplace mechanism, that can is used for numerical values, and hence in the location privacy context.
It relies on adding random noise, whose magnitude depends on the \textit{sensitivity} of the query function issued on the dataset.
Intuitively, the sensitivity of a query function quantifies the impact that the addition or removal of a single element of a dataset could have on the output of this function.

\begin{definition}
  Let $f$ be a function that takes a dataset as input and produces a vector of reals, i.e., $f: \mathcal{D} \longrightarrow \mathbb{R}^n, n \in \mathbb{N}$.
  Let $D_1$ and $D_2$ be two datasets differing on at most one element.
  The sensitivity of $f$ is noted $\Delta f$ and defined, for all such datasets $D_1$ and $D_2$, as:
  \[\Delta f = \max_{D_1,D_2} || f(D_1) - f(D_2) ||_1. \]
\end{definition}

The sensitivity is defined independently of the underlying data, and only depends on the function under consideration.
In particular, for queries that are counting records (such as $Q$ in our previous example), $\Delta Q = 1$ because the addition or removal of a single record affects the count result by increasing or decreasing its value by 1.
Then, the Laplace mechanism adds Laplacian noise with mean 0 and scale parameter $\Delta f/\epsilon$ to the query's result\footnote{Proof of this is provided in~\cite{Dwork06}.}.
Consequently, the $\epsilon$-differentially privacy version of $Q$ is defined as $\hat{Q}(D) = Q(D) + Y$, where $Y \sim \text{Lap}(1/\epsilon)$.
That way, computing $Q(D_2)-Q(D_1)$ does not automatically result in 1, because of the added Laplacian noise.
The Laplace mechanism is of course only suitable for queries producing numerical results; another method exists for categorical values~\cite{McSherry07}, but it is outside of the scope of this paper.

Differential privacy supports the composition of functions, and the potential information leakage resulting of this composition can be quantified.
In the general case, when applying $n$ randomized independent algorithms $\mathcal{K}_1,...,\mathcal{K}_n$ that provide $\epsilon_1,...,\epsilon_n$-differential privacy, any composition of those algorithms provides $(\Sigma_i \epsilon_i)$-differential privacy.
This is known as \textit{sequential composition}.
This protection model assumes that each analyst has a global privacy budget.
In an interactive mode (i.e., online LPPMs), each time she issues an $\epsilon$-differentially private query, his privacy budget is reduced by $\epsilon$.
Once the budget is totally consumed, all subsequent queries from this analyst should be rejected.
It models the fact that once an information is learnt, it cannot be forgotten.
In practice, determining this privacy budget and its instantiation (global, per user, etc.) remains largely an open question.
Recent works (e.g.,~\cite{Hsu14}) address this question.

Differential privacy has generated an important literature these last few years with new models and inter-model connections~\cite{DBLP:series/synthesis/2016Domingo-Ferrer}, as well as new techniques such as randomised response~\cite{DBLP:conf/edbt/0009WH16} and its combination with sampling~\cite{Krishnan:2016:IDA:2872427.2883026} which achieves zero-knowledge privacy~\cite{Gehrke:2011:TPS:1987260.1987294} (a privacy bound tighter than differential privacy).

%%%%%%%%%%%%%%%%%%%%%%%%%%%%%%%%%%%%%%%%%%%%%
\subsection{Privacy metrics}
\label{sec:background:privacy}
%%%%%%%%%%%%%%%%%%%%%%%%%%%%%%%%%%%%%%%%%%%%%
To quantify the level of protection offered by an LPPM, we identify three categories of privacy metrics.

\begin{itemize}
  \item \textit{Formal guarantee} metrics adopt a theoretical approach to quantify the effect of an LPPM on mobility data.
  They use a well-defined and unambiguous framework to guarantee that a protected dataset has a certain level of privacy.
  As of now, there are two such guarantees commonly offered by LPPMs: $k$-anonymity and differential privacy (cf. Section~\ref{sec:background:notions}).
  \textit{$k$-anonymity}, applied to location privacy, states that during a given time window and inside a given area, there should be at least $k$ users.
  LPPMs then take different approaches to enforce this guarantee, for example by allowing users to specify the size of these areas or time windows as parameters, or by automatically adjusting them, such as they contain $k$ users.
  \textit{$\epsilon$-differential privacy} has been instantiated differently by different LPPMs.
  Usually, instead of protecting the presence or absence of individual users, as it is the case with classical differential privacy, LPPMs attempt to protect the presence or absence of individual locations.
  Hence, the goal is not anymore to hide that a user is part of a dataset, but to hide where she went.
  \item \textit{Data distortion} metrics compare privacy-related properties of mobility data before and after applying an LPPM on it.
  Indeed, using an LPPM is expected to hide sensitive information that was otherwise possible to obtain from actual mobility data.
  Examples of such metrics include computing the entropy of protected data or evaluating whether POIs can still be retrieved.
  \item \textit{Attack correctness} metrics evaluate the impact of a location privacy attack that could be ran by an adversary in order to gain knowledge about users (see Section~\ref{sec:threats} for the list of potential attacks).
  Shokri et al.~\cite{Shokri11} did an extensive work on the usage of attacks to quantify location privacy.
  They distinguish between three axes when evaluating the effectiveness of an attack: certainty, accuracy and correctness.
  Certainty is about the ambiguity of the attack's result; for example there is some uncertainty if a re-identification attack outputs three possible users, while the uncertainty is null if the same attack outputs a single user (independently of whether it is the correct answer).
  Accuracy is about taking into account that the attacker does not have unlimited computational resources; consequently, the output of his attack may be only an approximate response, e.g., by only taking into account a sample of all data at his disposal.
  Correctness quantifies the distance between the attack's result and the truth; it is what actually quantifies location privacy.
  An LPPM is expected to mitigate privacy attacks and lower (or even suppress) their harmful effects.
  As opposed to data distortion metrics, attack correctness metrics do not compare the effect of an attack before and after applying an LPPM, but rather evaluate directly the attack on a protected dataset, and use the actual dataset as ground truth to evaluate whether the attack was successful.
\end{itemize}

Very recently, a survey has specifically focused on reviewing and discussing privacy metrics~\cite{DBLP:journals/corr/WagnerE15}.

%%%%%%%%%%%%%%%%%%%%%%%%%%%%%%%%%%%%%%%%%%%%%
\subsection{Utility metrics}
\label{sec:background:utility}
%%%%%%%%%%%%%%%%%%%%%%%%%%%%%%%%%%%%%%%%%%%%%
To evaluate the quality of protected mobility data, we identify two categories of utility metrics.

\begin{itemize}
  \item \textit{Data distortion} metrics compare utility-related properties of mobility before and after applying an LPPM on it.
  Indeed, we expect that the LPPM will not distort all properties of a dataset and make it unusable.
  Examples of such metrics include evaluating the spatial and temporal imprecision and comparing the covered area.
  It is of purpose that we name this category the same way as for privacy metrics, because they do represent the same thing, but applied on different properties (privacy- or utility-related).
  If we go even further, it happens that some data distortion metrics are used one time as a privacy metric and the other time as a utility metric\footnote{A common example is a metric whose goal is to compare the distance between actual locations and protected locations. It can be viewed either as a privacy metric, because by distorting locations we hide where users were, or as a utility metric, if the LBS that we use or the task that the analyst wants to run requires spatial precision.}.
  \item \textit{Task distortion} metrics compare the result of some practical task on the data before and after applying an LPPM.
  For instance, these metrics can be interested in data mining tasks or analytics queries.
\end{itemize}

\begin{table*}[ht]
  \centering
  \renewcommand{\arraystretch}{1.3}
  \caption{Common datasets of mobility traces.}
  \label{tab:evaluation:datasets}
  \begin{tabular}{|l|l|l|l|l|}
    \hline
    \textbf{Dataset} & \textbf{Location} & \textbf{Time span} & \textbf{\#users} & \textbf{\#events} \\ 
    \hline
    Cabspotting & San Francisco, USA & 1 month & 536 & 11 million \\
    \hline
    MDC & Geneva, Switzerland & 3 years & 185 & 11 million \\
    \hline 
    Privamov & Lyon, France & 15 months & 100 & 156 million \\
    \hline
    Geolife & Beijing, China & 5,5 years & 178 & 25 million  \\ 
    \hline 
    T-Drive & Beijing, China & 1 week & 10,357 & 15 million  \\ 
    \hline
    Brightkite & Worldwide & 1.5 years & 58,228 & 4 million \\
    \hline
    Gowalla & Worldwide & 1.5 years & 196,591 & 6 million \\
    \hline
  \end{tabular}
\end{table*}

%%%%%%%%%%%%%%%%%%%%%%%%%%%%%%%%%%%%%%%%%%%%%
\subsection{Performance metrics}
\label{sec:background:performance}
%%%%%%%%%%%%%%%%%%%%%%%%%%%%%%%%%%%%%%%%%%%%%
Protecting mobility data can be resources greedy. To evaluate the performance an LPPM, four categories of metrics are commonly used.

\begin{itemize}
  \item \textit{Execution time} is a simple quantification of the time it takes for an LPPM to protect data.
  Of course, it does not have the same impact for real-time use cases, where a response is expected in a very short time frame (a few milliseconds, a few seconds at most), than for batch or offline use cases that do not expect an immediate answer.
  However, even for the latter, it is of importance as computational resources have a cost ("time is money").
  This execution time can be measured in various ways, for example in seconds or in CPU cycles.
  \item \textit{Communication overhead} metrics quantify the negative impact of applying an LPPM on the quantity of information that will be produced and exchanged through the network in online use cases.
  For online use cases, some LPPMs need to exchange more messages, or more answers are received from the LBS.
  Obviously, it has an impact on the execution time, but it can be measured separately.
  For offline use cases it is related to the size of the protected dataset;
  if bigger or more complex that the actual one, it can slow down the job of analysts and affect their experience when working with the dataset.
  \item \textit{Energy overhead} metrics measure the negative impact on the battery lifetime implied by using a given LPPM, when running it as an application on a mobile device.
  It is important to be quantified because it impacts the usability and adoption by end users.
  It is only applicable to online LPPMs.
  \item \textit{Scalability} measures how well an LPPM can face a high workload.
  For online LPPMs, scalability metrics are mostly related to the capability of handling a high volume of concurrent requests, while for offline LPPMs it concerns the ability to deal with datasets of large sizes.
\end{itemize}

%%%%%%%%%%%%%%%%%%%%%%%%%%%%%%%%%%%%%%%%%%%%%
\subsection{Trade-off between utility and privacy}
\label{sec:evaluation:tradeoff}
%%%%%%%%%%%%%%%%%%%%%%%%%%%%%%%%%%%%%%%%%%%%%
Protecting mobility data by an LPPM improves the privacy but impacts the quality of the resulting data: this is the trade-off between privacy and utility. 
The more the information is altered (e.g., modified or deleted), the less the protected data may be exploited.
The configuration of this trade-off (i.e., defining the levels of privacy and utility required for the protected data) closely depends on the considered use case.
For instance, a weather application requires a less precise location (it can accommodate with only a city name) than a navigation application (it needs to know in which street a users is located).
Consequently, both the privacy and the utility evaluations should be tailored to fit the actual use case.

LPPMs are usually configured through various configuration parameters, which greatly impact the resulting privacy and utility, and it is an uneasy task to correctly define LPPM configuration parameters.
As an example, Wait for Me~\cite{Abul10} takes at least five parameters, with some labeled as the "initial maximum radius used in clustering" or the "global maximum trash size".
While there are useful to fine-tune the behaviour of the algorithm, we do not expect final users to read the paper to understand what the trash is or how the clustering works.
Even the single $\epsilon$ parameter of geo-indistinguishability~\cite{Andres13} is tricky to configure, because it is expressed in meters$^{-1}$ and its impact is exponential.
Similarly, it is difficult for a final user who knows (usually) nothing about differential privacy to set it appropriately.

Recent works have been conducted to select and configure the best LPPM to use according to a set of objectives set by the user in term of utility and privacy.
For instance, ALP~\cite{Mapomme16a} relies on a greedy approach that iteratively evaluates the privacy and utility, thus refining the values of configuration parameters at each step.
While ALP can be used for online and offline LPPMs, PULP~\cite{pulp} proposes a framework allowing to automatically choose and configure offline LPPMs.
To do that, PULP explores and models the dependency that exists, for different LPPMs, between their configuration parameters and the privacy/utility metrics that one wants to maximise.

\begin{figure*}[htbp]
  \centering
  \small
  \subfloat{\includegraphics[height=0.07\textwidth]{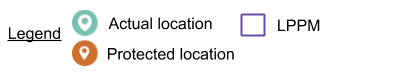}}
  \vspace{-6mm}
\end{figure*}

\begin{figure*}[htbp]
  \centering
  \small
  \subfloat[Trusted and Not Trusted Third Party (TTP up, NTTP bottom) -- Online]{\label{fig:ttp}\includegraphics[height=0.22\textwidth]{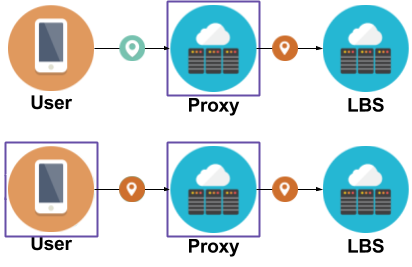}}
  \hspace{15mm}
  \subfloat[Peer-to-Peer (P2P) -- Online]{\label{fig:p2p}\includegraphics[height=0.22\textwidth]{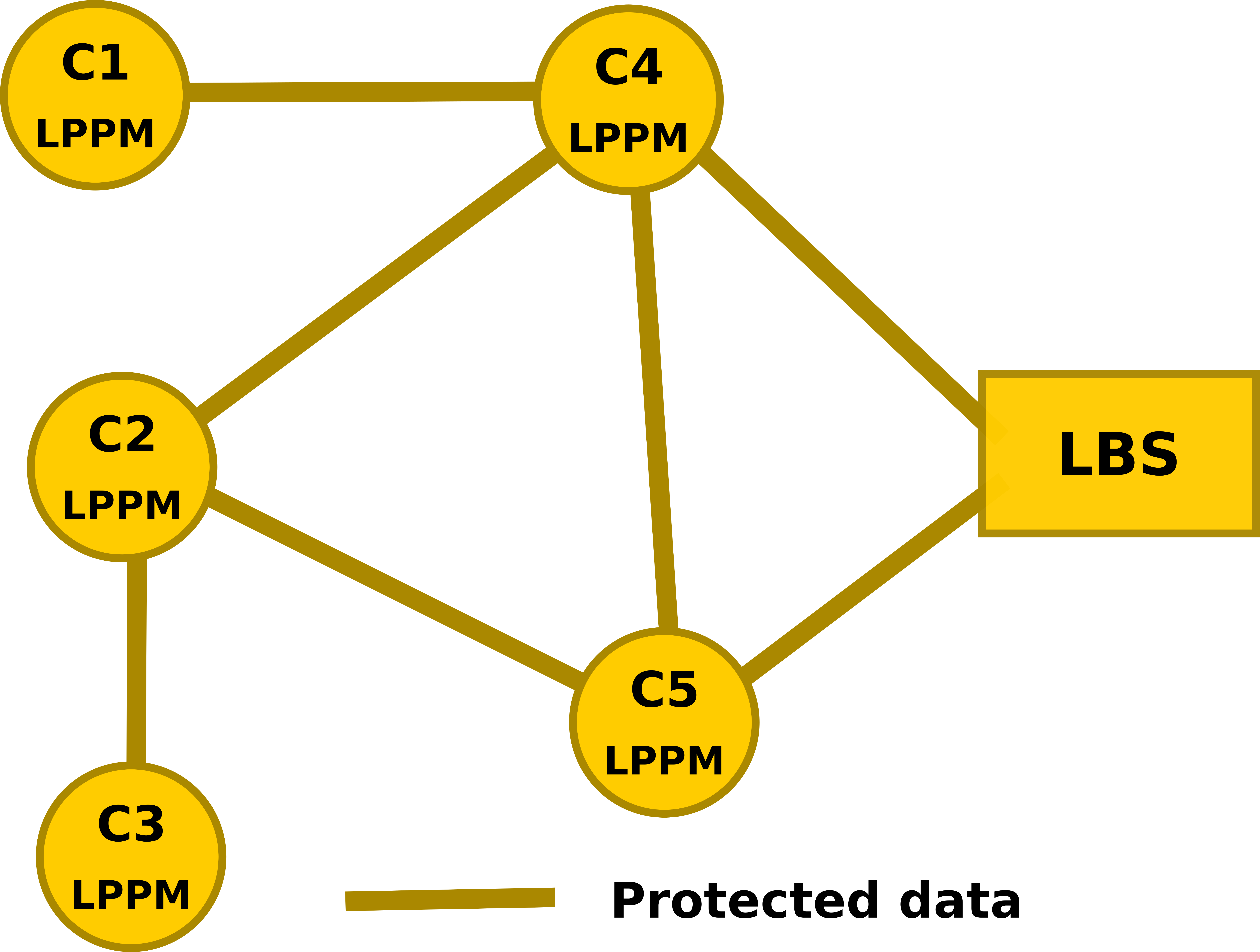}}
  \hspace{15mm}
  \subfloat[Local -- Online or Offline]{\label{fig:local}\includegraphics[height=0.22\textwidth]{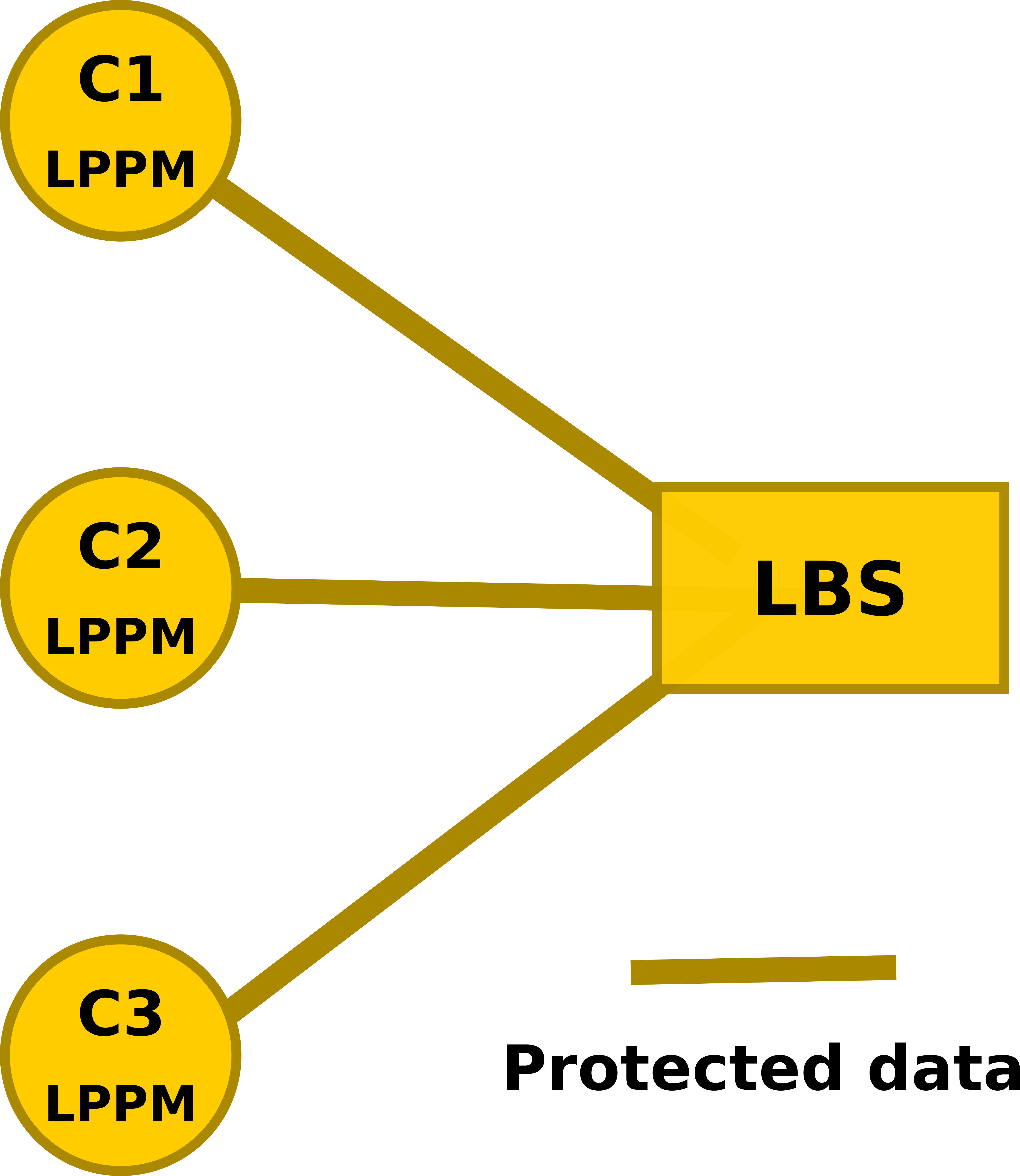}}
  \caption{Four different architectures can be adopted by LPPMs: Trusted Third Party (TTP), Non-Trusted Third Party (NTTP), Peer-to-Peer (P2P), and Local.}
\label{fig:architectures}

\end{figure*}

%%%%%%%%%%%%%%%%%%%%%%%%%%%%%%%%%%%%%%%%%%%%%
\subsection{Mobility datasets}%
\label{sec:evaluation:datasets}
%%%%%%%%%%%%%%%%%%%%%%%%%%%%%%%%%%%%%%%%%%%%%
To evaluate and compare protection mechanisms, we need to assess their effectiveness on mobility datasets, preferably including real mobility data of real users.
To this end, several initiatives have been conducted to publicly provide datasets coming from real-life data collections.
Table~\ref{tab:evaluation:datasets} lists some of the most common datasets used in the literature.

The Cabspotting dataset~\cite{cab} contains GPS traces of taxi cabs in San Francisco (USA), collected in May 2008. 
The Geolife dataset~\cite{Zheng09} gathers GPS trajectories collected from April 2007 to August 2012 in Beijing (China).
The MDC dataset~\cite{Laurila13,Kiukkonen10} involves 182 volunteers equipped with
smartphones running a data collection software in the Lake Geneva region (Switzerland), collected between 2099 and 2011.
A privacy protection scheme based on $k$-anonymity has been performed on the raw data before releasing the MDC dataset. 
As described in~\cite{Laurila13}, this privacy preserving operation includes many manual operations which have obviously an impact on the outcome of LPPMs, but these impacts are difficult to fully understand.
It includes not only locations coming from the GPS sensor, but also data from various other sensors (e.g., accelerometer, battery).
The Privamov dataset also gathers mobility data from multiple sensors (i.e., accelerometer, WiFi, cellular network).
This data collection took place from October 2014 to January 2016 and involves 100 students and staff from various campuses in the city of Lyon equipped with smartphones~\cite{benmokhtar:hal-01578557}.
T-Drive~\cite{Yuan10,Yuan11} is another dataset collected in Beijing and featuring taxi drivers.
It features a high number of users (more than 10,000) over a very short period of time (one week).

Other datasets come from geolocated social networks, rather than from a custom data collection campaign ran by academics.
These social networks allowed users to leave "check-ins" to places where they went, thus allowing to build sparse mobility traces for these users.
Two datasets are available in this category, coming from the (now closed) Brightkite and Gowalla~\cite{Cho11,snap} social networks.
They contain 4 million (respectively 6 million) check-ins collected between February 2009 and October 2010.
These datasets present a different kind of mobility data along with relationships between users in the network.

Lastly, another approach is to use \textit{synthetic datasets}, i.e., randomly generated mobility datasets.
Such generators include \textit{BerlinMOD}~\cite{Duntgen09}, Brinkhoff's generator~\cite{Brinkhoff02} and Hermoupolis~\cite{Pelekis15}.
Because they are generated, these datasets may not always model very realistically the human mobility and all the hazards attached to it.
However, this approach allows to use datasets of any size, for example to assess the scalability of algorithms to large-scale datasets.
Indeed, very few researchers have access to extremely large datasets (e.g., the dataset with 1.5 million individuals used by De Montjoye et al.~\cite{DeMontjoye13} or the dataset of 2.1 million individuals used by Wang et al. ~\cite{Wang18}).

Past experimental results have demonstrated that datasets can have an important impact on the evaluation of an LPPM.
LPPMs providing $k$-anonymity are among the most sensitive to this aspect.
For example, it is far easier to likely provide $k$-anonymity with an important value of $k$ with a large dataset than a small one.
Similarly, the sparsity of datasets will also be of high importance in $k$-anonymous LPPMs, because this protection scheme is particularly sensitive to the co-location of users (i.e., users being at the same place at the same time).
From one evaluation~\cite{Abul10} to another~\cite{Mapomme15b} of the same LPPM, results can indeed largely vary, due to the considered datasets and their associated features.
In the former case, the dataset is generated using Brinkhoff's generator, and represents one day of mobility with 4.7 million events.
In the latter case, the real-life datasets Geolife, Cabspotting and MDC are used, which contain between 11 and 25 million events, but scattered across several weeks or even years.
Another factor, besides the size of a datasets, is the kind of users that it contains.
Indeed, the taxi drivers inside the Cabspotting dataset will likely have a different behaviour that the ordinary people who are part of the MDC dataset.

\section{Architectures of LPPMs}
\label{sec:archi}

In addition to their different use cases, LPPMs can leverage four different architectures.
These architectures are depicted in Figure~\ref{fig:architectures}.
The local architecture is used by both online and offline LPPMs, while the Trusted Third Party (TTP), the Non-Trusted Third Party (NTTP), and Peer-to-Peer (P2P) architectures are only used by online LPPMs.

\begin{itemize}
  \item The \textit{TTP} architecture requires a trusted third party proxy server.
  It means there is an external entity that has access to the actual data coming from all users.
  \item The \textit{NTTP} architecture still involves one or several third party servers but they do not need to be trusted. However, the LPPM is designed in such a manner that this third party cannot represent a privacy threat, even if malicious or colluding with the LBS. In this scheme, the behaviour of the LPPM is usually split and implemented on both the client and the proxy server.
  \item The \textit{P2P} architecture requires no external server, but it requires users devices taking part in the system to exchange information in a peer-to-peer fashion in order to protect their data.
  Such LPPMs engage users devices in a collaborative privacy protocol before they send their data to an LBS.
   \item The \textit{local} architecture does not require any communication with another party to protect data.
  LPPMs entirely autonomous and process everything locally, on the device on which they are executed (i.e., a users devices or a server operated by the LBS).
  They may need access to external databases, in which case the latter are expected to be entirely available locally.
\end{itemize}

%% HERE IS THE TABLE OF ONLINE LPPMS.
\begin{table*}[th!]
\centering
\renewcommand{\arraystretch}{1.1}
\caption{List of \textit{online} LPPMs studied in this survey, with metrics used by their authors to evaluate them.}
\label{tab:online_lppms}
\vspace{4mm}

\begin{tabular}{|l|c||c|c|c|c|c||c|c|c||c|c|c|c|}
\hline
& & \multicolumn{5}{c||}{Privacy} & \multicolumn{3}{c||}{Utility} & \multicolumn{4}{c|}{Performance} \\
\hline
Protection mechanism & \rotatebox{90}{Architecture} & % 2 cols
  \rotatebox{90}{Differential privacy} & \rotatebox{90}{$k$-anonymity} & \rotatebox{90}{Attack correctness} & \rotatebox{90}{Data distortion} & \rotatebox{90}{Ad-hoc metric} & % 5 cols
  \rotatebox{90}{Data distortion} & \rotatebox{90}{Query distortion} & \rotatebox{90}{Ad-hoc metric} & % 3 cols
  \rotatebox{90}{Execution time} & \rotatebox{90}{Communication cost} & \rotatebox{90}{Energy overhead} & \rotatebox{90}{Scalability} \\ % 4 cols
\hline
\multicolumn{14}{c}{\textbf{Mix-zones}} \\
\hline
Beresford et al.~\cite{Beresford04} & TTP &
  & & \OK & & &
  & & & 
  \OK & & & \\
Freudiger et al.~\cite{Freudiger09} & TTP &
  & & \OK & & &
  & & \OK &
  & & & \\
Traffic-aware mix-zones \cite{Liu12} & TTP &
  & & \OK & & &
  & & \OK &
  \OK & & & \\
\textit{MobMix}~\cite{Palanisamy11} & TTP &
  & & \OK & & &
  & & &
  \OK & & & \OK \\
Gong et al.~\cite{Gong17} & TTP/P2P &
  & & & & &
  & & \OK &
  & & & \OK \\
\hline
\multicolumn{14}{c}{\textbf{Generalization-based mechanisms}} \\
\hline
\textit{CliqueCloak}~\cite{Gedik05} & TTP & 
  & \OK & & & &
  & \OK & &
  \OK & \OK & & \\
\textit{Casper}~\cite{Mokbel06} & TTP &
  & \OK & & & &
  & & &
  \OK & & & \\
\textit{P2P cloaking}~\cite{Chow06} & P2P &
  & \OK & & & &
  & \OK & &
  \OK & \OK & & \\
\textit{PRIV\'{E}}~\cite{Ghinita07} & P2P &
  & \OK & & & \OK &
  & & &
  \OK & \OK & & \OK \\
\textit{PrivacyGrid}~\cite{Bamba08} & TTP &
  & \OK & & & &
  & \OK & &
  \OK & & & \OK \\
Xu and Cai~\cite{Xu09} & TTP &
  & & & & \OK &
  & & \OK &
  \OK & & \OK & \OK \\
Agir et al.~\cite{Agir14} & Local &
  & & & \OK & \OK &
  \OK & & &
  & & & \\
Ngo et al.~\cite{Ngo15} & Local &
  \OK & & & \OK & &
  & & \OK &
  \OK & & & \\
\textit{ReverseCloak}~\cite{Li15} & TTP &
  & \OK & \OK & & \OK &
  & & &
  \OK & & & \\
Huguenin et al.~\cite{Huguenin17} & Local &
  & & & & &
  & \OK & \OK &
  & & & \\
\hline
\multicolumn{14}{c}{\textbf{Dummies-based mechanisms}} \\
\hline
Realistic fake trips~\cite{Krumm09a} & Local &
  & & & & &
  & & &
  & & & \\
Synthetic fake trips~\cite{Bindschaedler16} & Local &
  & & \OK & & &
  & \OK & \OK &
  & \OK & & \\
Kido et al.~\cite{Kido05} & Local &
  & \OK & \OK & & &
  & & &
  & & & \\
You et al.~\cite{You07} & Local &
  & \OK & \OK & & &
  & & &
  & & & \\
\textit{MobiPriv}~\cite{Stenneth10} & TTP &
  & \OK & & \OK & &
  & \OK & &
  \OK & & & \\
\textit{SpotME}~\cite{Quercia11} & Local &
  & \OK & & & &
  & \OK & &
  \OK & & & \\
Kato et al.~\cite{Kato12} & Local &
  & \OK & & \OK & \OK &
  & & &
  & & & \\
\textit{SybilQuery}~\cite{Shankar09} & Local &
  & \OK & & & \OK & 
  & & &
  \OK & \OK & & \\
\hline
\multicolumn{14}{c}{\textbf{Perturbation-based mechanisms}} \\
\hline
\textit{Geo-indistinguishability}~\cite{Andres13} & Local &
  \OK & & & & &
  & \OK & &
  & \OK & & \\
Path cloaking~\cite{Hoh07} & TTP &
  & & & \OK & &
  \OK & &
  & & & & \\
\textit{CAP}~\cite{Pingley09} & Local &
  & \OK & & & &
  & \OK & &
  \OK & & & \\
Temporal clustering~\cite{Assam11} & TTP &
  & \OK & & & &
  \OK & \OK &
  & \OK & & & \\
Location truncation~\cite{Micinski13} & Local &
  & & & & &
  & \OK & &
  & & & \\
Bordenabe et al.~\cite{Bordenabe14} & Local &
  \OK & & \OK & & &
  \OK & & \OK &
  \OK & & & \\
Oya et al.~\cite{Oya17} & Local &
  \OK & & \OK & \OK & &
  \OK & & &
  & & & \\
Predictive geo-indistinguishability~\cite{Chatzikokolakis14} & Local &
  \OK & & & & &
  \OK & & \OK &
  & & & \\
Elastic geo-indistinguishability~\cite{Chatzikokolakis15} & Local &
  \OK & & & \OK & &
  \OK & & &
  & & & \\
\textit{LocLok}~\cite{Xiao17} & Local &
  \OK & & & & &
  & & &
  & & & \\
\textit{PIVE}~\cite{Yu17} & Local &
  \OK & & \OK & & \OK &
  \OK & & &
  & & & \\
\hline
\multicolumn{14}{c}{\textbf{Protocol-based mechanisms}} \\
\hline
\textit{Louis, Lester and Pierre}~\cite{Zhong07} & P2P &
  & & & & &
  & & &
  \OK & & & \\
\textit{PrivStats}~\cite{Popa11} & P2P &
  & & & & &
  & \OK & &
  \OK & & & \OK \\
\textit{MobiCrowd}~\cite{Shokri11b} & P2P &
  & & & & \OK &
  & & &
  & & & \\
\textit{C-Hide\&Seek, C-Hide\&Hash}~\cite{Mascetti11} & NTTP &
  & & & & &
  & \OK & &
  \OK & \OK & & \\
\textit{SRide}~\cite{Aivodji18} & NTTP &
  & & & & &
  & & &
  \OK & \OK & & \OK \\
PIR~\cite{Ghinita08} & NTTP &
  & & & & &
  & & &
  \OK & & & \\
\textit{Trust No One}~\cite{Jaiswal10} & NTTP &
  & & & & &
  & & &
  & & & \\
Narayanan et al.~\cite{Narayanan11} & P2P/TTP &
  & & & & &
  & & &
  \OK & & & \\
\textit{Koi}~\cite{Guha12} & NTTP &
  & & & & &
  & & &
  & & & \OK \\
\textit{Zerosquare}~\cite{Pidcock13} & NTTP &
  & & & & &
  & & &
  \OK & & & \\
Outsourced garbled circuit~\cite{Carter13} & NTTP &
  & & & & &
  & & &
  \OK & \OK & & \OK \\
\hline
\multicolumn{14}{c}{\textbf{Rule-based mechanisms}} \\
\hline
\textit{ipShield}~\cite{Chakraborty14} & Local &
  & & & & &
  & & &
  \OK & & \OK & \\
\textit{LP-Guardian}~\cite{Fawaz14} & Local &
  & & & & &
  \OK & & &
  \OK & & \OK & \\
\hline
\end{tabular}
\end{table*}

%% HERE IS THE TABLE OF OFFLINE LPPMS.
\begin{table*}[t]
\centering
\renewcommand{\arraystretch}{1.2}
\caption{List of \textit{offline} LPPMs studied in this survey, with metrics used by their authors to evaluate them.}
\label{tab:offline_lppms}
\vspace{4mm}

\begin{tabular}{|l||c|c|c|c|c||c|c|c||c|c|}
\hline
& \multicolumn{5}{c||}{Privacy} & \multicolumn{3}{c||}{Utility} & \multicolumn{2}{c|}{Perf.} \\	
\hline
Protection mechanism & % 1 cols
  \rotatebox{90}{Differential privacy} & \rotatebox{90}{$k$-anonymity} & \rotatebox{90}{Attack correctness} & \rotatebox{90}{Data distortion} & \rotatebox{90}{Ad-hoc metric} & % 5 cols
  \rotatebox{90}{Data distortion} & \rotatebox{90}{Query distortion} & \rotatebox{90}{Ad-hoc metric} & % 3 cols
  \rotatebox{90}{Execution time} & \rotatebox{90}{Scalability} \\ % 2 cols
\hline
\multicolumn{11}{c}{\textbf{Generalization-based mechanisms}} \\
\hline
Nergiz et al.~\cite{Nergiz08} &
  \OK & & \OK & & &
  & & \OK &
  \OK & \\  
\textit{Never Walk Alone}~\cite{Abul08} &
  & \OK & & \OK & &
  & \OK & &
  \OK & \\
\textit{Wait for Me}~\cite{Abul10} &
  & \OK & & \OK & &
  \OK & \OK & &
  \OK & \OK \\
Yarovoy et al.~\cite{Yarovoy09} &
  & \OK & & & &
  \OK & \OK & &
  & \\
Differentially private grids~\cite{Qardaji13} &
  \OK & & & & &
  & \OK & &
  & \\
\textit{GLOVE}~\cite{Gramaglia15} &
  & \OK & & & &
  \OK & & &
  & \\
Gramaglia et al.~\cite{Gramaglia17} &
  & \OK & & & &
  \OK & & \OK &
  & \\
\hline
\multicolumn{11}{c}{\textbf{Dummies-based mechanisms}} \\
\hline
Realistic fake trips~\cite{Krumm09a} &
  & & & & &
  & & &
  & \\
Synthetic fake trips~\cite{Bindschaedler16} &
  & & \OK & & &
  & \OK & \OK &
  & \\
\textit{Hermes++}~\cite{Pelekis11} &
  & \OK & & & &
  & \OK & \OK & \OK
  & \\
\hline
\multicolumn{11}{c}{\textbf{Perturbation-based mechanisms}} \\
\hline
\textit{Geo-indistinguishability}~\cite{Andres13} &
  \OK & & & & &
  & \OK & & & \\
Path confusion~\cite{Hoh05} &
  & & \OK & & \OK &
  \OK & & &
  & \\
\textit{PINQ}~\cite{McSherry09} &
  \OK & & & & &
  & & &
  & \\
Chen et al.~\cite{Chen12} &
  \OK & & & & &
  & \OK & &
  \OK & \OK \\
Jiang et al.~\cite{Jiang13} &
  \OK & & & & &
  \OK & & &
  & \\
\textit{DP-WHERE}~\cite{Mir13} &
  \OK & & & & &
  \OK & & &
  & \\
Acs et al.~\cite{Acs14} &
  \OK & & & & &
  \OK & & &
  & \\
Riboni et al.~\cite{Riboni14} &
  \OK & \OK & & & &
  \OK & \OK & &
  & \\
\textit{Promesse}~\cite{Mapomme15b} &
& & & \OK & &
\OK & \OK & &
\OK & \OK \\
\hline
\end{tabular}
\end{table*}

\section{Location privacy preservation}
\label{sec:lppms}

In order to mitigate location privacy threats, LPPMs have been introduced.
Their goal is to transform mobility data in order to protect it and prevent threats such as the ones presented in Section~\ref{sec:threats}.
As presented in Section~\ref{sec:introduction}, we distinguish between two main use cases of LPPMs: \textit{online} and \textit{offline}.
In the online case, the LPPM protects either on-the-fly or by batch the mobility data before it even reaches the LBS.
In the offline case, the LPPM is applied on an entire dataset before its publication.

In this section, we survey existing works about LPPMs, and organise them into six categories.
We summarise this categorisation in Table~\ref{tab:online_lppms} and Table~\ref{tab:offline_lppms} for online and offline LPPMs, respectively.
Moreover, we indicate for each LPPM its architecture and the categories of metrics that were used to evaluate it by its authors.
For the sake of completeness, we distinguish in these tables between differential privacy and k-anonymity for privacy formal guarantees, and mention when an ad-hoc metric was used to evaluate LPPMs.
Ad-hoc metrics encompass metrics that do not fit in our classification, usually because they measure something that is unique to the considered LPPM (e.g., something related to its algorithm and that cannot be made generic to all LPPMs).

%%%%%%%%%%%%%%%%%%%%%%%%%%
\subsection{Mix-zones}
\label{sec:lppms:mixzones}
%%%%%%%%%%%%%%%%%%%%%%%%%%
Mix-zones are a concept introduced by Beresford and Stajano.~\cite{Beresford03}, taking its roots in the seminal work of Chaum~\cite{Chaum81} about mix networks.
The mix-zones model applies to mobile users communicating with LBSs, by using a pseudonym instead of their real identity (e.g., real name, IP or MAC address).
In this context, a mix-zone is an area where movements of users are not tracked, and consequently where users cannot communicate with an LBS.
When a user leaves a mix-zone, she receives a new pseudonym chosen among the pseudonyms of users inside the mix-zone.
It means that when $k$ users are inside a mix-zone at the same time, their identities will be shuffled, providing some of $k$-anonymity and resulting in an attacker's confusion.

\subsubsection{Online mechanisms}
The initial model was further refined by Beresford and Stajano~\cite{Beresford04}, providing a more formal mathematical model as well as a location privacy metric, from the point of view of the attacker.
A question that arises with mix-zones is where to place them.
This has been tackled by Freudiger et al.~\cite{Freudiger09}, with the goal to maximise location privacy while taking into account the negative impact of mix-zones on utility.
Practically, they use a new metric called mobility profiles to theoretically compute the effectiveness of a mix-zone, and then solve the mix-zones placement problem as an optimisation problem.
Another solution to the mix-zones placement problem was proposed by Liu et al.~\cite{Liu12}.
They model the city as a graph, where nodes are venues (i.e., places of interest inside a city such as monuments, restaurants, cinemas, etc.) and the road network is used to create edges connecting those venues.
On one hand, an LBS can have side information on this graph and use it to re-identify users.
On the other hand, information about traffic is used to compute the optimal placement of mix-zones as an optimisation problem.
\textit{MobiMix} proposed by Palanisamy and Liu~\cite{Palanisamy11} is another solution leveraging the road network for optimising the mix-zones placement.
They consider the speed of users as a side channel that could be used to re-identify them.
They also propose different manners to construct mix-zones, designed to defeat timing attacks.
Gong et al.~\cite{Gong17} proposed a socially-aware way to exchange pseudonyms.
They model the decision of changing a pseudonym as a game, which takes into account social ties between users.

\subsubsection{Discussion}
Overall, mix-zones require an important number of users to be effective. %suffer of a non-negligible weakness which is that they need "a lot" of users to be effective.
Indeed, if too few users participate to the system, it is not very likely that they will meet at any time during the day.
We believe that this critical mass of users is too important to make mix-zones usable for individual users willing to protect their privacy in online use cases.
Moreover, mix-zones need to rely on a third party to provide the pseudonyms, and a trusted third party to handle the swapping of pseudonyms.
Introducing another trusted party, in place of the LBS, does not appear to be a desirable property.

%%%%%%%%%%%%%%%%%%%%%%%%%%%%%%%%%%%%%%%%%%%%
\subsection{Generalization-based mechanisms}
\label{sec:lppms:cloaking}
%%%%%%%%%%%%%%%%%%%%%%%%%%%%%%%%%%%%%%%%%%%%
Generalization methods have been successfully applied to provide $k$-anonymity in the location privacy context, through the concept of spatial cloaking introduced by Gruteser et al.~\cite{Gruteser03}.
Broadly speaking, the idea is to report an coarser information instead of the exact location of users.
Besides the effect of reducing the precision of the information, this also allows to create cloaking areas in which at least $k$ users are at any given moment.

\subsubsection{Online mechanisms}
Gedik and Liu introduced \textit{CliqueCloak}~\cite{Gedik05}, which is a system generating cloaking areas on-the-fly, as messages arrive, before sending them to an LBS.
Users specify a value for $k$, the maximum size of the cloaking area and the maximum time between the engine transmits the message to the LBS.
Because the engine needs enough messages to enforce $k$-anonymity, messages can be delayed or cancelled if there are not enough queries coming from the same area.
\textit{Casper}~\cite{Mokbel06} is a spatial cloaking architecture proposed by Mokbel et al.
It uses a location anonymizer (i.e., a trusted third party) which knows locations of all users and their privacy parameters (a $k$ parameter and a minimal cloaking area).
When a user sends his query to the anonymizer, the latter transforms it into a cloaked query and forwards it to the query processor.
The query processor needs to be able to understand such cloaked queries.
The response is then sent to the anonymizer, which refines it by using the actual user location and sends the final response to the user.
Chow et al. proposed \textit{P2P cloaking}~\cite{Chow06}, which is essentially an improvement of Casper.
Like in Casper, users specify a privacy profile with a $k$ parameter and a minimal cloaking area size.
However, instead of using a central trusted anonymizer, a peer-to-peer protocol enables nearby peers to generate cloaking areas.
Clients can then send themselves the query including the cloaking area, instead of their exact location, to an LBS.
\textit{PRIV\'{E}}~\cite{Ghinita07} is a decentralised architecture proposed by Ghinita et al.
Instead of having a central server building cloaking areas, a hierarchical and distributed index is used.
This index defines groups of at least $k$ users in the same vicinity (i.e., cloaking areas).
\textit{PrivacyGrid}~\cite{Bamba08} is a solution introduced by Bamba et al. focusing on speed and effectiveness.
It is an anonymisation proxy providing $k$-anonymity and $\ell$-diversity thanks to a dynamic grid cloaking algorithm.
It allows users to specify individually their requirements in terms of both privacy and utility.
However, this LPPM can fail to deliver a query to an LBS; the probability of a query to be effectively protected can be increased by allowing to delay its sending.
Xu and Cai~\cite{Xu09} proposed a "feeling-based" approach to location privacy.
The main idea is that users specify their privacy requirements by defining a public region that they would feel comfortable to be reported as their location.
The goal is to replace traditional models in which users have to specify parameters, such as the $k$ of $k$-anonymous LPPMs, which may not be very expressive to them.
The authors then propose a solution, based on a trusted server, to build cloaking areas matching such a privacy requirement of each user.
Li and Palanisamy introduced \textit{ReverseCloak}~\cite{Li15}, an LPPM providing $k$-anonymity for users moving inside a road network.
This approach differs from previous work in that it provides a multi-level reversible privacy model.
It means that different LBSs may have different access levels, and thus access more or less granular information.

Agir et. al~\cite{Agir14} introduced an adaptive mechanism to dynamically change the size of obfuscated areas hiding the exact location of users.
More precisely, the proposed solution locally evaluates the privacy level and enlarges the area until a target privacy level is achieved, or the information is too distorted (in which case the location cannot be released).
Ngo and Kim~\cite{Ngo15} proposed a protection mechanism trying to optimise the average size of cloaked areas generated by Hilbert curve methods.
They define a new privacy metric for cloaking areas, relying on $\epsilon$-differential privacy.
They explore a notion of identifiability, quantifying the probability of an attacker to identify the user's location from the cloaking area, that help them to choose a value for the $\epsilon$ parameter.
Huguenin et al.~\cite{Huguenin17} studied the effect of using generalization-based LPPMs while using LBSs such as Foursquare~\cite{foursquare} to leave check-ins.
Towards this end, they used a predictive model to quantify the effect of generalization on perceived utility.
They were able to predict with a small error the loss of utility caused by the generalization protection, allowing to implement efficiently LPPMs featuring both good privacy and utility properties.

\subsubsection{Offline mechanisms}
Nergiz et al. \cite{Nergiz08} proposed an algorithm to segment trajectories into groups of points providing $k$-anonymity.
Then, they introduced a randomised reconstruction algorithm that uses the cloaked areas obtained from the previous step to recreate trajectories, giving a protected version of the original trajectories.
Abul et al. proposed \textit{Never Walk Alone}~\cite{Abul08}, whose idea is to guarantee that at every instant there is at least $k$ users walking at a given distance of the others, thus creating cylinders within which users move.
This radius exploits the inherent incertitude that comes with GPS measurements to avoid distorting the data too much.
This mechanism has been later improved by \textit{Wait4Me}~\cite{Abul10}, which is more generic with respect to the input dataset it can protect, and scales better to large datasets.

When people move, they essentially move from one place to another, which are often POIs.
The list of these places can been considered as a quasi-identifier, which can be protected with a $k$-anonymity guarantee.
Yarovoy et al.~\cite{Yarovoy09} tackled the problem of creating optimal anonymisation groups for moving objects, which unlike traditional databases may not be disjoint.
They consider an attacker model where the latter is building an attack graph giving relationships between objects in the protected database and their identities in the raw database.
Qardaji et al.~\cite{Qardaji13} studied the usage of grids to partition the continuous space into a discrete domain.
They started with static grids and proposed a method to choose the grid's size, and then proposed a new approach using an adaptive grid.
Gramaglia and Fiore~\cite{Gramaglia15} proposed a measure of the anonymisability of mobility datasets, based on their spatio-temporal similarity.
Then, they introduced \textit{GLOVE}, an adaptive protection mechanism providing $k$-anonymity while reducing the utility loss.
It iteratively merges mobility traces that are the most similar, with respect to the previously introduced metric, until all groups of traces contain at least $k$ users.
Gramaglia et al.~\cite{Gramaglia17} introduced $k^{\tau,\epsilon}$-anonymity, which is a model extending $k$-anonymity to include temporal information.
In this setup, an attacker may have some background knowledge covering a continuous period of time of at most $\tau$, and is allowed to discover the mobility records of a targeted user for a period of at most $\epsilon$ (disjoint from $\tau$).
The authors propose a way to reach $k^{\tau,\epsilon}$-anonymity by relying on a spatio-temporal generalization.

\subsubsection{Discussion}
Overall, generalization-based LPPMs have the advantage to propose an easy to understand privacy model (e.g., $k$-anonymity).
This category of LPPMs is most adapted to offline scenarios with deterministic approaches.
Indeed, providing different protected versions of the same dataset with an non-deterministic approach can reveal additional information at each release.
Moreover, as outlined with \textit{Wait for Me}, these approaches faces to a scalability issue when the size of the dataset increases.

In online use cases, this protection scheme suffers from the same weakness than mix-zones, specifically the requirement to have enough users to be effective.
In addition, generalization-based LPPMs usually do not work with GPS coordinates but with areas or trajectory which is not immediately usable by existing LSBs.

%%%%%%%%%%%%%%%%%%%%%%%%%%%%%%%%%%%%%
\subsection{Dummies-based mechanisms}
\label{sec:lppms:dummies}
%%%%%%%%%%%%%%%%%%%%%%%%%%%%%%%%%%%%%
Instead of relying on other users to be hidden among them and obtain $k$-anonymity, as with generalization-based approaches, it is possible to generate fake users, called dummies.
In this scheme, the attacker may be aware that there are dummies inside the data it got, but the challenge here is to generate realistic fake data, ideally indistinguishable from the real data.

\subsubsection{Online mechanisms}
Dummies-based protection mechanisms are mostly used for online usage in the literature.
The basic idea is for each user to send multiple queries to an LBS, instead of a single one.
One of those queries contain his actual location, while the others contain fake locations.
Consequently, the LBS is not able to determine exactly where the user really is located.
Kido et al.~\cite{Kido05} were the first to introduce a protection mechanism using dummies.
They simply split the space into regions of a fixed size and generate dummies in neighbouring regions.
You et al.~\cite{You07} proposed another method to create fake trajectories.
They generate endpoints randomly and then generate trajectories between these new endpoints with two methods.
A first method generates a trajectory randomly by using vertical, horizontal and diagonal random speeds.
A second method intends to force intersections between trajectories of dummies and the real user's trajectory.
In that case, a dummy trajectory is obtained by rotating the real trajectory around a given point.
Stenneth et al.\cite{Stenneth10} presented \textit{MobiPriv}, which uses an anonymisation proxy through which all queries transit before being sent to an LBS.
Similarly to centralized protection mechanisms presented in Section~\ref{sec:lppms:cloaking}, the proxy of MobiPriv ensures k-anonymity by generating realistic looking dummies.
It also leverages a history of previous queries to prevent attacks using the intersection of multiple queries' results to infer new knowledge.
\textit{SpotME}, proposed by Quercia et al.~\cite{Quercia11}, also generates dummies within discrete regions.
However, this solution is specific to one use case: counting users inside regions.
Users report to be or not within a region according to a randomized mechanism.
The LBS can then "reverse" this mechanism and compute how many users are really within a region with a high probability.
If users are honest, there is at most an error of 11~\% in the final result.
Lato et al.~\cite{Kato12} presented a method to generate dummies that acknowledge the fact that users make stops during their mobility, while previous work often consider generating fake trajectories between two endpoints.
They assume user's movements are known in advance and use this knowledge to reduce user's traceability (i.e., increasing the confusion of an LBS about which of the dummies are the real users).
More specifically, each time the user stops, there is a possibility that a dummy can stop at the same place, thus creating a crossing between multiple paths and increasing the confusion.

Shankar et al. introduced \textit{SybilQuery}~\cite{Shankar09}, a solution to generate real-looking fake trips, especially suited for navigation applications.
Knowing the real trajectory that a user will do, SybilQuery will generate fake trips starting from and ending to different locations, but preserving properties such as the length of the trip and the semantics of the areas where endpoints are located (e.g., residential vs business areas).
When a user moves and sends queries to an LBS to get directions, fake users will also move along fake trips.
Krumm~\cite{Krumm09a} proposed a probabilistic model to generate fake driving trips.
Endpoints are chosen according to some probability model.
A route planner is used to generate a trajectory between trips, with some randomness injected into trajectory to prevent the optimal path to be always selected (indeed, users do not always follow the best path when driving).
Speeds are also drawn from a probability model, and some noise is finally added to each point to simulate GPS noise.
Bindschaedler et Shokri~\cite{Bindschaedler16} presented a way to generate synthetic mobility traces that share statistical features with real traces in a privacy-preserving way.
These synthetic traces are designed to be used instead of the real traces, thus presumably leaking no sensitive information.
They build a mobility model for each trace and an aggregate probabilistic mobility model about the entire dataset, and use it to synthetise fake traces from these models; which must satisfy a privacy test before being released.

\subsubsection{Offline mechanisms}
In an offline context, dummies can also be used to provide artificial $k$-anonymity when data would otherwise be discarded because there is no other user with a similar behaviour, although it is less used because it introduces an obvious error for analysts.
Pelekis et al. proposed \textit{Hermes++}~\cite{Pelekis11}, which is a privacy-preserving query engine.
It relies on the injection of dummies in query results, these dummies being designed to follow the behaviour of actual users.
This engine also has an auditing module that is able to detect if a sequence of queries can be harmful for the privacy of individuals (i.e., if trying to track users over time).

\subsubsection{Discussion}
The main issue with dummies-based LPPMs is their ability to produce real-looking dummies.
Indeed, a study of Peddinti et al.~\cite{Peddinti11} showed that \textit{SybilQuery} is very vulnerable to attacks based on machine learning.
They developed an algorithm able to correlate traces, and tested it against a dataset containing data of 85 taxi drivers around San Francisco.
SybilQuery was configured with $k=5$, which means that each mobility event generated by a driver was hidden among four other dummy events.
In the case of an attacker having access to a previous mobility dataset (i.e., forming the training dataset), their algorithm re-identified 93~\% of the users.
Furthermore, some of these algorithms (e.g.,~\cite{Shankar09,Krumm09a}) use an extensive amount of external knowledge, such as a graph modeling the road network, a route planner or census statistics about the population.
However, using and processing important external knowledges in online use cases represents a limiting factor for a usage on mobile device with limited computational and storage capacities.

%%%%%%%%%%%%%%%%%%%%%%%%%%%%%%%%%%%%%%%%%%
\subsection{Perturbation-based mechanisms}
\label{sec:lppms:perturbation}
%%%%%%%%%%%%%%%%%%%%%%%%%%%%%%%%%%%%%%%%%%
While protection mechanisms providing $k$-anonymity try to hide a user inside a crowd, other mechanisms rely on the alteration of the data to be sent to an LBS to protect it.
In this case the challenge is to have a trade-off between privacy (i.e., the data needs to be distorted enough to be protected) and the utility (i.e., if the data is too distorted, results from the LBS will be unusable).
Most mechanisms work by adding some (often random) noise to the underlying raw data.

\subsubsection{Online mechanisms}
Hoh et al.~\cite{Hoh07} proposed a mechanism playing with the confusion of an attacker.
They designed a privacy metric called \textit{time-to-confusion}, quantifying the duration for which a given user can be tracked by an attacker.
They developed an LPPM using an third party server which aims to maximise this metric, in the context of traffic monitoring.
Pingley et al. presented \textit{CAP}~\cite{Pingley09}.
This solution protects two channels by which a curious LBS could obtain information: the location contained inside the query and the user's IP address.
For the latter an improved routing algorithm in the Tor anonymising network~\cite{Dingledine04} is used, while for the former Hilbert curves are leveraged to generate fake locations close to the real one.
Assam and Seidl~\cite{Assam11} proposed a mechanism enforcing $k$-anonymity through temporal clustering of streams of mobility data.
Micinski et al.~\cite{Micinski13} studied the effect of location truncation, i.e., reducing the precision of latitude/longitude coordinates by dropping decimals, on a nearby venues finder application on Android.
This is a simple protection mechanism that can still be effective to protect someone's exact location.

Differential privacy has been generalised for location privacy by Andres et al. under the notion of \textit{geo-indistinguishability}~\cite{Andres13}.
Geo-indistinguishability is a formal notion of location privacy that bounds the probability of two points to be reported locations of the same real location within a given radius.
Thus, a user can quantify the level of privacy she wants within a specific area.
Practically, it is done through the $\epsilon$ parameter (the lower $\epsilon$, the higher the noise), resulting in $\epsilon$-geo-indistinguishability.
Authors proposed a way to provide geo-indistinguishability by adding noise drawn from a planar Laplace distribution to a real location.
Bordenabe et al.~\cite{Bordenabe14} proposed a method to construct an LPPM enforcing geo-indistinguishability that maximises the utility.
To achieve that, they relied on linear programming techniques, and proposed a way to reduce the number of constraints, this improving the time required to build the LPPM.
Oya et al.~\cite{Oya17} also studied the design of optimal LPPMs, including geo-indistinguishability among others.
They argue that, besides metrics based on attack correctness, auxiliary metrics should be taken into account when evaluating an LPPM.
Consequently, they introduced two such metrics, showed that one single metric is not sufficient to assess the efficiency of an LPPM, and used these additional metrics to design a new LPPM.
Due to temporal correlations between a user's locations, differential privacy proposed in geo-indistinguishability can be problematic because each time a location is protected, some more privacy is lost.
In other words, protecting a trace of $n$ locations with $\epsilon$-geo-indistinguishability results at the end in $n\epsilon$-geo-indistinguishability.
To overcome this limitation, Chatzikokolakis et al. proposed a predictive mechanism~\cite{Chatzikokolakis14} providing geo-indistinguishably, using prediction to avoid spending too much budget for each location protection.
With two different ways of spending the privacy budget this gives a substantial improvement over the original geo-indistinguishable mechanism.
The same authors also proposed another extension of geo-indistinguishability~\cite{Chatzikokolakis15}, which leverages contextual information to calibrate the amount of noise applied to disturb the mobility traces.
They consider two levels of sensitivity if the user is located in an urban environment, where there is a high density of venues around, or in a sparse countryside area.

To account the temporal correlation of mobility traces, Xiao and Xiong~\cite{Xiao15} also proposed a mechanism based on differential privacy and a new measure to evaluate the sensitivity of each protected location.
The same authors later proposed \textit{LocLok}~\cite{Xiao17}, an LPPM taking into account temporal correlations via a hidden Markov model.
In a nutshell, their mechanism maintains a hidden Markov model of possible actual locations each time a protected location is released, and then generates protected locations via a differentially private method taking into account this Markov model.
Yu et al. introduced \textit{PIVE}~\cite{Yu17}, whose goal is to enforce at the same time two privacy guarantees: geo-indistinguishability and protection against adversary attacks.
Their solution works in two steps: first a protection location set is generated from the actual location and a minimum attacker error specified by the user.
Then a protected location, having differential privacy guarantees, is generated from the protection location set.

\subsubsection{Offline mechanisms}
Hoh and Gruteser~\cite{Hoh05} introduced the idea of path confusion as an LPPM.
The idea is to make closer users' paths to cross when they are close enough, to augment the confusion of an adversary about which path belong to which user.
They formulate and solve this problem as a constrained non-linear optimisation problem.
We proposed \textit{Promesse}~\cite{Mapomme15b}, which is specifically designed to hide the POIs of users.
It works on mobility traces by smoothing speed, i.e., making the speed appear as being constant.
Therefore, users seem to be always moving, thus making it more difficult to guess were they stopped and what their POIs are.

\textit{PINQ} (for Privacy INtegrated Queries)~\cite{McSherry09} is an analytics platform allowing to execute queries against a data source while preserving privacy through differential privacy.
The data analyst writes his queries, specifies privacy budget $\epsilon$ that can be consumed, and the platform automatically takes care of returning results satisfying differentially private guarantees.
One of the proposed examples illustrates geo-located queries and shows that PINQ can be successfully applied in this context.
Chen et al.~\cite{Chen12} protected public transportation usage data, which can be seen for each user as a sequence of places (metro/bus stations) she went to.
They built a method to protect such data in a differentially private way and evaluated their mechanism by studying the impact of their protection on range queries and sequential pattern mining.
Differential privacy has been used by Jiang et al.~\cite{Jiang13} to protect ships' trajectories.
Endpoints of trajectories are preserved while intermediate locations are altered by adding some noise satisfying differential privacy guarantees.
\textit{DP-WHERE}~\cite{Mir13} is a method introduced by Mir et al. to generate synthetic Call Detail Records (CDRs) in a differentially private way.
They start by building a model of real CDRs, formed of several histograms, and then add noise to each of them to achieve differential privacy.
A synthetic CDR can be generated by using the private versions of the histograms.
Authors of geo-indistinguishability~\cite{Andres13} also presented an offline usage of their protection mechanism.
Acs et al.~\cite{Acs14} proposed a mechanism to protect spatio-temporal densities datasets, which reports counts of active users within small areas for given time windows.
Such data can be obtained for instance from call data records that are gathered by mobile phone operators.
The authors proposed an approach that adapts to the original data in order to guarantee differential privacy with the highest possible utility.
Counting users is one interesting thing to do with mobility data, but we want to publish entire trajectories and allow more mining tasks to be performed.
Riboni and Bettini~\cite{Riboni14} introduced a way to publish check-in data (e.g., from Swarm~\cite{swarm}) in a differentially private manner, with the goal to allow venue recommandation from this data.
They start by filtering check-ins that fall within regions of fixed size where a single user did too many check-ins, thus indicating that it may be an important or sensitive venue for him.
Then, some noise is added to the number of check-ins at each venue to enforce differential privacy, before to release these statistical values.

\subsubsection{Discussion}
Online perturbation-based LPPMs have the advantage to be working on a local architecture, i.e., they do not need an external trusted party, as opposed to generalization-based LPPMs which very often require one.
A large number of them also rely on differential privacy which can be performed locally.
However, an open question with this privacy model in an interactive mode (i.e., online use cases) represents the management of the privacy budget $\epsilon$ (e.g.,~\cite{Tang2017}).
In addition, as the meaning of $\epsilon$ is less intuitive than the meaning of $k$ (in $k$-anonymity), choosing the proper value of this parameter may also be less clear from a user point of view.
Lastly, some perturbation-based LPPMs do not rely on any formal privacy guarantee, which makes their guarantees harder to justify in practice.

%%%%%%%%%%%%%%%%%%%%%%%%%%%%%%%%%%%%%%
\subsection{Protocol-based mechanisms}
\label{sec:lppms:protocol}
%%%%%%%%%%%%%%%%%%%%%%%%%%%%%%%%%%%%%%
Protection mechanisms falling in previous categories rely on the alteration of mobility data in order to protect information.
The solutions adopted in this category is to propose protocols which preserve privacy by design.
These protocol-based mechanisms are generally more specific (e.g., getting nearby friends, counting people) but can achieve the best privacy guarantees.
They largely rely on encryption schemes offering strong privacy guaranties for specific use cases.

\subsubsection{Online mechanisms}
\textit{Louis, Lester and Pierre} are three protocols proposed by Zhong et al.~\cite{Zhong07} that can be used to locate nearby friends in a privacy-preserving way.
They are multi-parties protocols: a user, say Alice, initiates a communication with another user, say Bob, and tries to learn if she is within a given radius $r$ from herself.
Depending on the protocol, Alice learns nothing, Bob's exact location, Bob's exact distance or Bob's grid cell distance.
On the other side Bob learns nothing, Alice's exact location and/or the radius $r$.
Popa et al.~\cite{Popa11} introduced \textit{PrivStats}, a system that can be used to collect location-based aggregate statistics within defined geographic areas.
Users collaborate to send pre-aggregated and encrypted data to the LBS, which allows to hide the number of tuples and the time at which they were collected.
The LBS receives a constant number of (encrypted) values at fixed time intervals, combines them by using homomorphic encryption and asks a user to decrypt the final aggregate value.
Authors also propose a privacy-preserving accountability protocol without any trusted party to prevent clients from cheating.
\textit{MobiCrowd}~\cite{Shokri11b} is a protection mechanism designed by Shokri et al. relying on collaboration between users to avoid querying an LBS if the information is already available on another nearby user's device.
Each requested piece of information is stored inside a local buffer on users' devices and are given an expiration time.
Users issuing queries first broadcast it to neighbours in an attempt to get the answer without contacting the LBS.
Mascetti et al.~\cite{Mascetti11} proposed a protocol for proximity notifications of nearby friends in a space divided into cells.
Two protocols are introduced in the article.
With \textit{C-Hide\&Seek}, each user sends to an LBS her current encrypted cell identifier, the encryption key being shared between each pair of friends.
This way, users can learn in which cell their friends are, even if they are not nearby.
With \textit{C-Hide\&Hash}, each user sends a hash of her salted location to the LBS, the salt being the key shared between each pair of friends.
By computing the hashes of cells around and running a private set intersection protocol with the LBS, users can learn which friends are nearby, but not their distance.
Narayanan et al.~\cite{Narayanan11} introduced another protocol for testing the proximity of friends.
They transformed the problem from proximity testing to equality testing, and then presented two protocols for private equality testing, one using a peer-to-peer architecture and the other relying on a trusted server.
They finally introduced another solution relying on nearby location tags (e.g., WiFi broadcast packets) detected by users, who may then run a private set intersection protocol to infer their proximity.
\textit{SRide}~\cite{Aivodji18} is a privacy-preserving ridesharing system proposed by A\"{\i}vodji et al.
The goal of such a system is to prevent the ridesharing platform to learn sensitive information about the origin and destination of the users' trips.
They leverage tools such as homomorphic encryption and secure multiparty computation; the ridesharing platform is still needed but do not have to be trusted anymore.
Overall, the SRide protocol can be executed in 5 to 9 seconds, with a communication overhead comprised between 3 and 6 MB.

\textit{Private information retrieval} (PIR), first theorised by Chor et al.~\cite{Chor95}, is a schema allowing someone to retrieve a row from a database without letting it know what she wants to retrieve.
Ghinita et al.~\cite{Ghinita08} proposed to apply PIR to N-nearest neighbours spatial queries, that can be used for example to look for nearby venues (i.e., restaurants, monuments, etc.).
They introduced a way to index spatial information in a PIR-compliant way by using Hilbert space-filling curves.
\textit{Trust No One}~\cite{Jaiswal10}, proposed by Jaiswal and Nandi, uses \textit{pseudonymised locations} to represent locations by an identifier without revealing actual coordinates and \textit{pseudonymised identifiers} to represent entities by an identifier.
These two pseudonyms are generated by two different entities, respectively a mobile operator and an LBS.
Finally, a decentralized matching service, that does not know anything about location or identity of entities, has the responsibility to answer queries.
\textit{Koi}~\cite{Guha12} is a platform proposed by Guha et al.
It relies on two non-colluding servers, namely the \textit{matcher} and the \textit{combiner}.
The matcher knows about entities and locations but nothing about links between them (i.e., which location belongs to which entity).
The combiner knows the mapping between entities and locations but nothing about actual content of these entities and locations.
A communication protocol between the matcher and the combiner allows to answer queries by performing a privacy-preserving matching.
Instead of directly querying Koi, mobile devices set up triggers reacting to some events (e.g., getting notified when there is a restaurant at less than 500 meters around me).
Application developers must hence create event-centric applications instead of location-centric applications.
Pidcock and Hengartner proposed \textit{Zerosquare}~\cite{Pidcock13}, which relies on two non-colluding servers, one of them storing a user-indexed database and the other a location-indexed database.
Moreover, some cloud components owned by service providers can be allowed individually by each user to access data contained in the location-indexed database.
This is the mobile device itself that queries the two databases and join information coming from each of them.
Garbled circuits where theorised by Yao~\cite{Yao82} and allow two parts to privately evaluate the result of a generic function.
Carter et al.~\cite{Carter13} proposed a way to outsource the evaluation of such garbled circuits.
Since they require a high computational power, outsourcing their evaluation in the cloud allows to speed up the processing and let a mobile device use garbled circuit despite a low computational capacity.
The challenge is to preserve privacy guarantees even with an untrusted cloud.
As an example they implemented a privacy-preserving navigation application that mainly consists in a Dijkstra shortest-path algorithm used to privately get directions between two (private) points while taking into account (private) hazards that can occur along the path.

\subsubsection{Discussion}
Overall, protocol-based LPPMs exhibit the best privacy and utility trade-off.
This is because they are tailored to answer more specific use cases, instead of trying to solve a large range of use cases.
But it also implies that such approaches may be too specific, e.g., \cite{Narayanan11,Mascetti11} are designed only to detect nearby friends.
They also do not interact with existing LBSs, because they essentially replace them.
While this gives to the LPPM designer much more freedom, it may seriously slow down their adoption, as new infrastructures need to be deployed to support them.
Moreover, because of the cost of the cryptographic primitives, such approaches may still be practically unusable because of their algorithmic complexity which impacts the execution time.
For example, finding navigation directions using outsourced garbled circuits~\cite{Carter13} takes about 15 minutes, and for a road network graph composed of only 100 vertices.

%%%%%%%%%%%%%%%%%%%%%%%%%%%%%%%%%%
\subsection{Rule-based mechanisms}
\label{sec:lppms:rules}
%%%%%%%%%%%%%%%%%%%%%%%%%%%%%%%%%%
Some believe that one-size-fits-all protection mechanisms are unrealistic.
This is why some protection mechanisms implement various state-of-the-art solutions and follow a set of rules to decide of the most appropriate countermeasure to take in the current situation.

\subsubsection{Online mechanisms}
Chakraborty et al. proposed \textit{ipShield}~\cite{Chakraborty14}, which is a framework, implemented on Android, leveraging a rules engine to protect location privacy.
Users define which threats they want to be protected against, with a priority level.
The system then leverages a database of inference attacks to recommend protection rules to apply on each sensor (i.e., not only the GPS but also the accelerometer, the gyroscope, etc.).
Users can also define their own rules, using contextual information and specifying actions to take on sensor data.
\textit{LP-Guardian}~\cite{Fawaz14} is a software running on Android proposed by Fawaz and Shin to protect location privacy of Android smartphones users.
They designed a framework to protect privacy against different threats: tracking threat, identification threat and profiling threat.
It uses a decision tree to decide which action to perform in a given situation, leveraging the context (e.g., application being used, location) and a combination of statically-defined and user-defined rules.

\subsubsection{Discussion}
As rule-based mechanisms essentially rely on a composition of LPPMs from other categories, they inherit the associated pros and cons.
Rule-based mechanisms can cover a wider range of use cases, however the side effects of these compositions can also jeopardise the privacy protections.

\section{Open challenges}
\label{sec:challenges}

Although the literature in the location privacy field is quite large, there are still several open challenges.
We present in this section some of them based on the lessons we have learned from our experience.
The first two challenges are more research-oriented, while the three last challenges are more technical

%%%%%%%%%%%%%%%%%%%%%%%%%%%%%%%%%%%%%%%%%
\subsection{Quantifying location privacy}
%%%%%%%%%%%%%%%%%%%%%%%%%%%%%%%%%%%%%%%%%
Evaluating the efficiency of a protection mechanism is not an easy task.
However, as it appears by looking at Table~\ref{tab:online_lppms} and Table~\ref{tab:offline_lppms}, that there is a high heterogeneity when it comes to the metrics used to evaluate LPPMs, although it is possible to categorise them in a small number of categories.
This makes it very difficult to fairly evaluate and compare LPPMs.
We believe that an important research direction is to be able to have a common framework to evaluate LPPMs, by using a set of well-defined and accepted metrics, across all three dimensions of privacy, utility and performance.

Few works have been done in this direction. 
Shokri et al.~\cite{Shokri11} were the first to propose a formal framework to evaluate the efficiency of a protection mechanism by using the impact of a location privacy attack.
They formally define three dimensions when evaluating privacy: the accuracy, the certainty and the correctness.
They introduced different attacks and used their framework to evaluate a few protection mechanisms.
It is worth noting that their framework is available as an open-source C++ tool~\cite{lpm}.
However, this work is only applicable to a subset of LPPMs, which fit into a probabilistic framework, and is only interested in evaluating privacy.
Later the ALP framework~\cite{Mapomme16a} (available as an open source tool~\cite{alp}) proposed a more generic support to configure LPPMs from a set of privacy and utility objectives.
However,  the number of available metrics and the definition of the objectives are still somewhat limited, and the convergence to appropriate configuration parameters is not ensured.

Those works both introduce a model and a number of metrics to evaluate privacy and utility of LPPMs.
Indeed, despite formal guarantees are needed in most contexts, they do not always translate well how an LPPM behaves in practice, as it has been shown in several works (e.g.,~\cite{Shokri10,Mapomme14,Oya17b}).
On the privacy side, we advocate (similarly to, e.g.,~\cite{Yu17}) that metrics relying on adversary attacks should be considered as complimentary to the formal guarantees.
The emergence of data anonymization and de-anonymization challenges~\cite{challenge} is promising to propose new protection schemes and assess existing ones.
On the utility side, besides the classical information theoretic metrics such as the entropy, there is a need to consider more application-driven use cases.

%%%%%%%%%%%%%%%%%%%%%%%%%%%%%%%%%%%%%%%%%%%%%%
\subsection{Towards new protection mechanisms}
%%%%%%%%%%%%%%%%%%%%%%%%%%%%%%%%%%%%%%%%%%%%%%
Differential privacy has met a large interest and generated an important literature since its introduction in 2006~\cite{DBLP:series/synthesis/2016Domingo-Ferrer,DBLP:conf/edbt/0009WH16}.
Several recent works on location privacy use and apply this approach for the protection of geolocated information (e.g.,~\cite{Riboni14,Chatzikokolakis14,Chen12}).
We believe that this privacy model is still promising and will continue to generate an important literature in the next few years.
As noticed in Section~\ref{sec:lppms:perturbation}, how to manage the $\epsilon$ privacy budget in an interactive mode still remains an important question.
But besides its wide adoption and interest in the research community, other guarantees are still worth being used such as $l$-diversity and $t$-closeness. 
Chatzikokolakis et al.~\cite{Chatzikokolakis17} explore more in depth the deterministic (e.g., $k$-anonymity) and non-deterministic (e.g., differential privacy) methods that can be used to design modern LPPMs.

Another promising track of research concerns the composition of LPPMs (e.g., combining $k$-anonymity with $l$-diversity where $l<k$).
We previously classified such approaches as rule-based LPPMs (Section~\ref{sec:lppms:rules}).
These approaches rely on different LPPMs and use one or another depending on the actual situation.
Because there is no one-size-fits-all LPPM, combining existing LPPMs allows to cover a larger variety of use cases.
However, quantifying the guarantee offered by the composition of heterogeneous LPPMs is challenging.

Yet another way to tackle the lack of a one-size-fits-all LPPM, instead of composing existing LPPMs, is to dynamically alter the level of protection offered by single LPPM.
Because not every data is equally sensitive and needs to be similarly protected, such approaches bring adaptivity.
Dynamically adapting the offered privacy level avoids over protecting data, and consequently provides a better utility.
This has been explored by some previous works (e.g,~\cite{Agir14,Chatzikokolakis15,Huguenin17,Mapomme16a}).
However, few of these works take into account the semantics of visited places.
For example, it may be more important to protect that a user went to a hospital than to protect that he is shopping inside a mall.
Providing adaptivity with respect to the semantics of some place (Section~\ref{sec:threats:semantics} for more on the semantics aspects) could be a smart way to provide users with a tailored LPPM.

Recently, we have seen the development of privacy-by-design approaches~\cite{Guha12,Pidcock13}.
Privacy-by-design has been theorised by the information and privacy commissioner of Ontario, Canada~\cite{pbd}.
In a nutshell, it relies on seven core principles: proactivity, privacy as the default setting, privacy embedded in the design, full functionality, end-to-end security, visibility/transparency and user-centricity.
In other words, it advocates for systems where privacy is integrated since the beginning as a requirement and by default, where the interests of the user come first, and without sacrificing the quality of service.
Despite seeming utopian, this goal is actually reachable as soon as we throw away the LBS stack as we know it today.
With privacy-by-design architectures, there is no need anymore to alter mobility data, as the LBS itself integrates privacy as a first class citizen.

%%%%%%%%%%%%%%%%%%%%%
\subsection{Datasets}
%%%%%%%%%%%%%%%%%%%%%
As shown in Section~\ref{sec:evaluation:datasets}, the research community has at its disposal a few real-life mobility datasets to evaluate its work.
However, despite, several initiatives that have been conducted to publicly provide datasets coming from real-life data collections, all these datasets remain small and involve a limited number of users.
This lack of large datasets strongly limit the ability of researchers to test their solutions under real condition.
Providing golden standards in terms of large mobility data collections is definitely appealing and would be very useful to better compare LPPMs.  
There is a real need to share methodologies and tools around those collections, and make them available to the research community.
Some efforts are already going into that direction, such as the Funf~\cite{Aharony11} and APISENSE~\cite{Haderer13} platforms, or the Crawdad~\cite{crawdad} community.
Lastly, open data initiatives followed by many organisations and cities (e.g., the city of Montreal provides trajectory data~\cite{opendatamontreal}) are also promising to provide open and large datasets.

%%%%%%%%%%%%%%%%%%%%%%%%%%%%
\subsection{Users awareness}
%%%%%%%%%%%%%%%%%%%%%%%%%%%%
Most of the users are not aware about the risk related to the exploitation of their mobility data.
There is a lack of tools to improve users' awareness on this.
To give an example, \textit{Please Rob Me}~\cite{pleaserobme} is a website whose goal is to "raise awareness about over-sharing", by showing it is possible to infer from geo-located tweets whether users are at home.
Moreover, people are not aware of the value of their mobility data, certainly because they do not know the amount of knowledge that can be derived from it.
A study showed that people would share their mobility trace in exchange of a little amount of money (the median was \textsterling10 or \textsterling20 for a commercial usage in~\cite{Danezis05}) or a gift (1~\% of chances to win a US\$200 MP3 player in~\cite{Krumm09b}).
We advocate it is one of the mission of researchers to raise awareness on societal problems such as privacy.
Besides talks targeted towards the general public, tools could be developed to highlight privacy issues and the benefits of using an LPPM.
\textit{FindYou}~\cite{Riederer16} is an example of such a tool that allowed users to visualise what could be inferred from the data collected by online LBSs such as Foursquare, Instagram or Twitter.
To go a step further, it would be very interesting to additionally show the impact of an LPPM on the collected data, and the benefits it brings to users' privacy.

%%%%%%%%%%%%%%%%%%%%%%%%%%%%%%%%%%
\subsection{Implementation effort}
%%%%%%%%%%%%%%%%%%%%%%%%%%%%%%%%%%
To be used, protection mechanisms obviously need to be implemented and made available.
Very few solutions are freely downloadable and usable without reimplementing them from scratch.
A notable work is \textit{ipShield}~\cite{Chakraborty14}, which is actually implemented on the Android platform (though not necessarily installable trivially by end-users, because it is tightly integrated in the Android kernel).
Geo-indistinguishability~\cite{Andres13} has also been implemented by its authors as a browser extension~\cite{lg} working with several popular browsers.
This extension easily allows users to benefit from some privacy when using geolocated services through their Web browser.
Another example is Aircloak~\cite{aircloak}, a project that aims to propose a trusted sensitive data collection architecture with privacy-preserving querying capabilities.
By using several layers of noise, as well as maintaining a history of previous queries, the application is able to detect combinations of queries that could result in a privacy leak and prevent them.
With ACCIO~\cite{Mapomme18}, we proposed an experimental platform to experiment with location privacy.
This platform implements several state-of-the-art mobility data manipulation routines, privacy and utility metrics, and LPPMs such as geo-indistinguishability~\cite{Andres13} or Wait4Me~\cite{Abul10}.

Lastly, we have also seen large companies taking steps to actually implement privacy-preserving measures in their products.
Apple for example, uses differential privacy for some machine learning applications~\cite{appledp}, such as the keyboard suggestions.
Google also successfully integrated RAPPOR~\cite{Erlingsson14} into its Chrome browser, allowing to get usage statistics in a privacy preserving way.
The latter also relies on differential privacy, although it requires a large number of users to behave properly (the experiments were conducted with 1 million users).

\section{Conclusion}
\label{sec:conclusion}
In this article, we surveyed the latest works about computational location privacy.
At the best of our knowledge, it is the first survey to propose a unified view on both online and offline protection mechanisms, and putting the evaluation metrics as first-class citizens.
This shows that online and offline protection mechanisms can be based on the same underlying primitives (e.g., differential privacy), while providing appropriate algorithms suited for the considered use case.
While literature is already rich in various protection mechanisms, we outlined the lack of standard methods to compare these mechanisms.

According to the title of this survey, there is still a long road until location privacy can be democratised, both politically (i.e., being accepted in the users' mind) and technically (i.e., having production-quality software).
But some recent theoretical and practical works are encouraging and show the way to what future research in location privacy could be.

\ifCLASSOPTIONcaptionsoff
  \newpage
\fi

\IEEEtriggeratref{152}
\bibliographystyle{IEEEtran}
\bibliography{IEEEabrv,bibli}

\end{document}